\documentclass[prd,nofootinbib,english,a4paper]{revtex4-2}
\usepackage{amsmath}
\usepackage{amsfonts}
\usepackage{amssymb}
\usepackage{amsthm}
\usepackage{accents}
\usepackage[utf8]{inputenc}
\usepackage[T1]{fontenc}
\usepackage[colorlinks=true]{hyperref}
\usepackage[final]{showlabels}
\allowdisplaybreaks

\newcommand{\lc}[1]{\accentset{\circ}{#1}}
\newcommand{\dd}{\mathrm{d}}
\newcommand{\ginv}[2]{\underaccent{\mathsf{#2}}{\hat{\boldsymbol{#1}}}}
\newcommand{\gdep}[2]{\underaccent{\mathsf{#2}}{\hat{#1}}}

\begin{document}

\title{Gauge-invariant cosmological perturbations in general teleparallel gravity}

\author{Lavinia Heisenberg}
\email{L.Heisenberg@ThPhys.Uni-Heidelberg.DE}
\affiliation{Institut für Theoretische Physik, Philosophenweg 16, 69120 Heidelberg, Germany}

\author{Manuel Hohmann}
\email{manuel.hohmann@ut.ee}
\affiliation{Laboratory of Theoretical Physics, Institute of Physics, University of Tartu, W. Ostwaldi 1, 50411 Tartu, Estonia}

\begin{abstract}
We study linear cosmological perturbations in the most general teleparallel gravity setting, where gravity is mediated by the torsion and nonmetricity of a flat connection alongside the metric. For a general linear perturbation of this geometry around a homogeneous and isotropic background geometry, we derive the irreducible decomposition of the perturbation variables, as well as their behavior under gauge transformations, i.e., infinitesimal diffeomorphisms generated by a vector field. In addition, we also study these properties for the most general set of matter variables and gravitational field equations. We then make use of these result to construct gauge-invariant perturbation variables, using a general approach based on gauge conditions. We further calculate these quantities also in the metric and symmetric teleparallel geometries, where nonmetricity or torsion is imposed to vanish. To illustrate our results, we derive the energy-momentum-hypermomentum conservation equations for both the cosmological background and the linear perturbations. As another example, we study the propagation of tensor perturbations in the $f(G)$, $f(T)$ and $f(Q)$ class of theories.
\end{abstract}

\maketitle

\section{Motivation}\label{sec:motivation}
Numerous open questions in gravity theory become apparent from observations in cosmology, such as the cosmic microwave background radiation~\cite{Aghanim:2019ame,Aghanim:2018eyx,Aghanim:2018oex,Akrami:2019izv,Akrami:2019bkn,Akrami:2018odb}, the large scale structure~\cite{Ahumada:2019vht,Alam:2020sor}, gravitational waves~\cite{Abbott:2016blz,TheLIGOScientific:2017qsa} and supernovae~\cite{Scolnic:2017caz}. In order to describe these observations, one needs to study the evolution of both the universe as a whole, modeled by a homogeneous and isotropic background geometry and matter distribution, as well as perturbations of this background. A thorough understanding of such cosmological perturbations and their dynamics imposed by the gravitational interaction is therefore an important necessity for describing and explaining the modern observations in cosmology.

Cosmological perturbations in gravity have been studied for a long time, starting with the case of (pseudo-)Riemannian spacetime geometry, which is employed by the standard formulation of general relativity and the most well-known class of its extensions, in which the gravitational interaction is attributed to the curvature of the metric-compatible, torsion-free Levi-Civita connection~\cite{Lifshitz:1945du,Lifshitz:1963ps,Hawking:1966qi,Harrison:1967zza}. This task is significantly simplified by the fact by understanding how perturbations transform under gauge transformations, i.e., infinitesimal diffeomorphisms which retain the nature of the spacetime geometry as a small perturbation of a cosmologically symmetric background. From these gauge transformations, one can derive a set of gauge-invariant perturbation variables, which describe the physical information contained in the metric perturbations as well as the perturbations of the matter variables, so that they become independent of the arbitrary gauge choice. The resulting gauge-invariant perturbation theory is one of the cornerstones of modern cosmology~\cite{Bardeen:1980kt,Kodama:1985bj,Mukhanov:1990me,Malik:2008im}.

Despite its overwhelming success in describing observations from laboratory scales up to galactic scales, general relativity is challenged by the aforementioned open questions, as well as the open question how it can be reconciled with quantum theory. This situation motivates the study of modified gravity theories~\cite{CANTATA:2021ktz}. While numerous theories depart from the standard formulation of general relativity in terms of the curvature of the Levi-Civita connection of a Riemannian spacetime, also other formulations in terms of the torsion or nonmetricity of a flat connection exist and can be used as potential starting points for the construction of modified gravity theories~\cite{BeltranJimenez:2019tjy,Hohmann:2022mlc}. Focusing on general relativity alone, one finds that these formulations are equivalent in the sense that they lead to field equations which possess the same solutions for the metric irrespective of the geometric properties of the connection under consideration. However, modifications thereof lead to essentially inequivalent classes of theories, whose field equations possess different solutions.

In order to make use of the cosmological perturbation theory also in teleparallel gravity theories, one needs to understand both the most general teleparallel background geometries and their linear perturbations. While the former have been thoroughly studied and completely classified~\cite{Hohmann:2019nat,Hohmann:2020zre,Hohmann:2021ast,Heisenberg:2022mbo}, perturbations have so far only been studied in the metric teleparallel framework, in which only torsion is present, but nonmetricity is imposed to vanish~\cite{Golovnev:2018wbh,Hohmann:2020vcv}. The aim of this article is to complete this study by providing a comprehensive overview of the perturbations of general teleparallel geometry, where both torsion and nonmetricity are present, as well as the two more restricted cases of metric and symmetric teleparallel geometries, which arise as particular sectors of the more general geometries. For this purpose, we derive the irreducible decomposition of the perturbations in their scalar, vector and tensor components, and study their behavior under gauge transformations. From these we derive a general framework for constructing sets of gauge-invariant field variables. While the main focus of this article is on the geometry, we also exemplify its use by studying the general form of the energy-momentum-hypermomentum conservation laws and the propagation of tensor perturbations in a simple class of teleparallel gravity theories.

The article is structured as follows. We start with a brief review of general teleparallel gravity in section~\ref{sec:gentele}, as well as the homogeneous and isotropic background in section~\ref{sec:background}. We then define the perturbation variables and their irreducible decomposition in section~\ref{sec:cosmotelepert}. Gauge transformations and gauge-invariant quantities are discussed in section~\ref{sec:gaugetrans}. As a potential application, we derive the energy-momentum-hypermomentum conservation in section~\ref{sec:enmomhypcons}. As another example, we derive the tensor propagation equations in section~\ref{sec:example}. We end with a conclusion in section~\ref{sec:conclusion}.

\section{General teleparallel gravity}\label{sec:gentele}
We start our discussion with a brief review of the different flavors of teleparallel gravity theories and their underlying geometry. Our conventions for the geometric quantities appearing in this article are laid out in section~\ref{ssec:gentele}. The general form of their first-order perturbations is discussed in section~\ref{ssec:telepert}. Finally, in section~\ref{ssec:genactfield} we review the general structure of the action and field equations of teleparallel gravity theories.

\subsection{General teleparallel geometry}\label{ssec:gentele}
We start by giving a brief review of the general teleparallel geometry, which constitutes the fundamental fields of general teleparallel gravity theories. As a subclass of metric-affine geometry, the dynamical fields are given by a Lorentzian metric \(g_{\mu\nu}\) and an affine connection with coefficients \(\Gamma^{\mu}{}_{\nu\rho}\), which is imposed to be flat,
\begin{equation}\label{eq:nocurv}
R^{\rho}{}_{\sigma\mu\nu} = \partial_{\mu}\Gamma^{\rho}{}_{\sigma\nu} - \partial_{\nu}\Gamma^{\rho}{}_{\sigma\mu} + \Gamma^{\rho}{}_{\lambda\mu}\Gamma^{\lambda}{}_{\sigma\nu} - \Gamma^{\rho}{}_{\lambda\nu}\Gamma^{\lambda}{}_{\sigma\mu} = 0\,.
\end{equation}
It is important to note that this connection is different from the Levi-Civita connection, whose coefficients are the Christoffel symbols
\begin{equation}
\lc{\Gamma}^{\mu}{}_{\nu\rho} = \frac{1}{2}g^{\mu\sigma}(\partial_{\nu}g_{\sigma\rho} + \partial_{\rho}g_{\nu\sigma} - \partial_{\sigma}g_{\nu\rho})\,,
\end{equation}
where we use an overset circle to denote any quantity related to the Levi-Civita connection. The difference of the coefficients of the two connections constitutes a tensor field, which can be written as
\begin{equation}\label{eq:conndec}
\Gamma^{\mu}{}_{\nu\rho} - \lc{\Gamma}^{\mu}{}_{\nu\rho} = M^{\mu}{}_{\nu\rho} = K^{\mu}{}_{\nu\rho} + L^{\mu}{}_{\nu\rho}\,,
\end{equation}
where the distortion \(M^{\mu}{}_{\nu\rho}\) is composed of the contortion
\begin{equation}\label{eq:contortion}
K^{\mu}{}_{\nu\rho} = \frac{1}{2}\left(T_{\nu}{}^{\mu}{}_{\rho} + T_{\rho}{}^{\mu}{}_{\nu} - T^{\mu}{}_{\nu\rho}\right)\,,
\end{equation}
as well as the disformation
\begin{equation}\label{eq:disformation}
L^{\mu}{}_{\nu\rho} = \frac{1}{2}\left(Q^{\mu}{}_{\nu\rho} - Q_{\nu}{}^{\mu}{}_{\rho} - Q_{\rho}{}^{\mu}{}_{\nu}\right)\,,
\end{equation}
and these are defined through the torsion
\begin{equation}\label{eq:torsion}
T^{\mu}{}_{\nu\rho} = \Gamma^{\mu}{}_{\rho\nu} - \Gamma^{\mu}{}_{\nu\rho}\,,
\end{equation}
and the nonmetricity
\begin{equation}\label{eq:nonmetricity}
Q_{\mu\nu\rho} = \nabla_{\mu}g_{\nu\rho} = \partial_{\mu}g_{\nu\rho} - \Gamma^{\sigma}{}_{\nu\mu}g_{\sigma\rho} - \Gamma^{\sigma}{}_{\rho\mu}g_{\nu\sigma}\,.
\end{equation}
In particular, it follows that the affine connection is fully characterized by its torsion and nonmetricity. We will make use of this fact when we specify the most general homogeneous and isotropic teleparallel geometry below. Note, however, that these two tensor fields cannot be chosen arbitrarily, but their values are restricted by the condition~\eqref{eq:nocurv} of vanishing curvature, which together with the decomposition~\eqref{eq:conndec} becomes
\begin{equation}\label{eq:curvdec}
R^{\mu}{}_{\nu\rho\sigma} = \lc{R}^{\mu}{}_{\nu\rho\sigma} + \lc{\nabla}_{\rho}M^{\mu}{}_{\nu\sigma} - \lc{\nabla}_{\sigma}M^{\mu}{}_{\nu\rho} + M^{\mu}{}_{\tau\rho}M^{\tau}{}_{\nu\sigma} - M^{\mu}{}_{\tau\sigma}M^{\tau}{}_{\nu\rho} \equiv 0\,.
\end{equation}
This relation is particularly useful, as it allows relating the curvature tensor \(\lc{R}^{\mu}{}_{\nu\rho\sigma}\) of the Levi-Civita connection to the distortion and its covariant derivative. We will make use of this relation when we discuss gravity theories in section~\ref{sec:example}.

\subsection{Perturbation of teleparallel geometry}\label{ssec:telepert}
We start by introducing a convenient notation for the perturbation of the fundamental fields, which we will use throughout the remainder of this section. For the metric perturbation we will write
\begin{equation}\label{eq:metricpert}
\delta g_{\mu\nu} = g_{\mu\nu} - \bar{g}_{\mu\nu} = \varsigma_{\mu\nu}\,,
\end{equation}
while we write the perturbation of the teleparallel affine connection as
\begin{equation}\label{eq:connpert}
\delta\Gamma^{\mu}{}_{\nu\rho} = \Gamma^{\mu}{}_{\nu\rho} - \bar{\Gamma}^{\mu}{}_{\nu\rho} = \bar{\nabla}_{\rho}\lambda^{\mu}{}_{\nu}\,.
\end{equation}
The latter is the most general linear perturbation of the affine connection which preserves its flatness. This can be seen by calculating the curvature perturbation
\begin{multline}
\delta R^{\mu}{}_{\nu\rho\sigma} = \bar{\nabla}_{\rho}\delta\Gamma^{\mu}{}_{\nu\sigma} - \bar{\nabla}_{\sigma}\delta\Gamma^{\mu}{}_{\nu\rho} + \bar{T}^{\tau}{}_{\rho\sigma}\delta\Gamma^{\mu}{}_{\nu\tau}\\
= \bar{\nabla}_{\rho}\bar{\nabla}_{\sigma}\lambda^{\mu}{}_{\nu} - \bar{\nabla}_{\sigma}\bar{\nabla}_{\rho}\lambda^{\mu}{}_{\nu} + \bar{T}^{\tau}{}_{\rho\sigma}\bar{\nabla}_{\tau}\lambda^{\mu}{}_{\nu} = \bar{R}^{\mu}{}_{\tau\rho\sigma}\lambda^{\tau}{}_{\nu} - \bar{R}^{\tau}{}_{\nu\rho\sigma}\lambda^{\mu}{}_{\tau} = 0\,,
\end{multline}
which vanishes, since the background curvature vanishes by assumption. Similarly, one calculates the perturbation of the torsion,
\begin{equation}
\delta T^{\mu}{}_{\nu\rho} = 2\delta\Gamma^{\mu}{}_{[\rho\nu]} = 2\bar{\nabla}_{[\nu}\lambda^{\mu}{}_{\rho]}\,,
\end{equation}
as well as the nonmetricity,
\begin{equation}
\delta Q_{\rho\mu\nu} = \bar{\nabla}_{\rho}\delta g_{\mu\nu} - \delta\Gamma^{\sigma}{}_{\mu\rho}\bar{g}_{\sigma\nu} - \delta\Gamma^{\sigma}{}_{\nu\rho}\bar{g}_{\mu\sigma} = \bar{\nabla}_{\rho}\varsigma_{\mu\nu} - \bar{g}_{\sigma\nu}\bar{\nabla}_{\rho}\lambda^{\sigma}{}_{\mu} - \bar{g}_{\mu\sigma}\bar{\nabla}_{\rho}\lambda^{\sigma}{}_{\nu}\,.
\end{equation}
From these formulas it is easy to derive a few special cases. In the symmetric teleparallel case, we impose vanishing torsion both for the background and the perturbation; this is achieved by choosing the connection perturbation to be of the form
\begin{equation}\label{eq:connperts}
\lambda^{\mu}{}_{\nu} = \bar{\nabla}_{\nu}\zeta^{\mu}\,.
\end{equation}
Similarly, for the metric teleparallel case, we have vanishing nonmetricity for the background, and so we can pull the metric under the derivative, contract and obtain the condition
\begin{equation}\label{eq:connpertm}
\varsigma_{\mu\nu} = 2\lambda_{(\mu\nu)}
\end{equation}
for the vanishing nonmetricity perturbation. In the following, we prefer to keep the metric perturbation as a fundamental variable, as its components will be connected directly to observables, and we split the connection perturbation \(\lambda_{\mu\nu}\) into a symmetric and antisymmetric part, so that we can replace the former with the metric perturbation when we impose vanishing nonmetricity.

\subsection{Generic action and field equations}\label{ssec:genactfield}
In order to derive a perturbative expansion of the field equations of teleparallel gravity theories around a cosmologically symmetric background, we start with a brief review of their general structure, and introduce the relevant notation. For this purpose we assume an action of the form
\begin{equation}
S[g, \Gamma, \chi] = S_{\text{g}}[g, \Gamma] + S_{\text{m}}[g, \Gamma, \chi]\,,
\end{equation}
where the gravitational part \(S_{\text{g}}\) of the action depends only on the metric and the connection, while the matter part \(S_{\text{m}}\) also depends on some set of matter fields \(\chi^I\), whose components we do not specify further and simply label them with an index \(I\). The variation of the gravitational part of the action with respect to the metric and the connection then takes the form~\cite{Hohmann:2021fpr,Hohmann:2022mlc}
\begin{equation}\label{eq:gravactvar}
\delta S_{\text{g}} = -\int_M\left(\frac{1}{2}W^{\mu\nu}\delta g_{\mu\nu} + Y_{\mu}{}^{\nu\rho}\delta\Gamma^{\mu}{}_{\nu\rho}\right)\sqrt{-g}\dd^4x\,,
\end{equation}
where we introduced the tensor fields \(W^{\mu\nu}\) and \(Y_{\mu}{}^{\nu\rho}\). For the matter action, the variation takes the form~\cite{Hehl:1994ue}
\begin{equation}\label{eq:matactvar}
\delta S_{\text{m}} = \int_M\left(\frac{1}{2}\Theta^{\mu\nu}\delta g_{\mu\nu} + H_{\mu}{}^{\nu\rho}\delta\Gamma^{\mu}{}_{\nu\rho} + U_I\delta\chi^I\right)\sqrt{-g}\dd^4x\,,
\end{equation}
where \(U_I = 0\) are the matter field equations, and we introduced the energy-momentum \(\Theta^{\mu\nu}\) and hypermomentum \(H_{\mu}{}^{\nu\rho}\). When deriving the gravitational equations from this action, it must be taken into account that the connection and hence also its variation is not arbitrary, but must satisfy the flatness condition~\eqref{eq:nocurv}. This can be achieved equivalently either by introducing Lagrange multipliers, or by performing a restricted variation~\cite{Hohmann:2021fpr}. Here we follow the latter, and write the variations in the form
\begin{equation}
\delta g_{\mu\nu} = \varsigma_{\mu\nu}\,, \quad
\delta\Gamma^{\mu}{}_{\nu\rho} = \nabla_{\rho}\lambda^{\mu}{}_{\nu}\,,
\end{equation}
which are formally identical to the perturbations~\eqref{eq:metricpert} and~\eqref{eq:connpert} introduced earlier, and hence preserve the flatness of the connection. In terms of these, the variation of the gravitational part of the action reads
\begin{equation}
\delta S_{\text{g}} = -\int_M\left(\frac{1}{2}W^{\mu\nu}\varsigma_{\mu\nu} + Y_{\mu}{}^{\nu\rho}\nabla_{\rho}\lambda^{\mu}{}_{\nu}\right)\sqrt{-g}\dd^4x = -\int_M\left[\frac{1}{2}W^{\mu\nu}\varsigma_{\mu\nu} + (M^{\omega}{}_{\rho\omega}Y_{\mu}{}^{\nu\rho} - \nabla_{\rho}Y_{\mu}{}^{\nu\rho})\lambda^{\mu}{}_{\nu}\right]\sqrt{-g}\dd^4x\,,
\end{equation}
and analogously for the matter part of the action, where we have performed integration by parts in the last step.

Performing either of these procedures, one finds the field equations~\cite{Hohmann:2022mlc}
\begin{equation}\label{eq:gentelefield}
W_{\mu\nu} = \Theta_{\mu\nu}\,, \quad
\nabla_{\rho}Y_{\mu}{}^{\nu\rho} - M^{\omega}{}_{\rho\omega}Y_{\mu}{}^{\nu\rho} = \nabla_{\rho}H_{\mu}{}^{\nu\rho} - M^{\omega}{}_{\rho\omega}H_{\mu}{}^{\nu\rho}\,.
\end{equation}
The second equation can also be rewritten by introducing the tensor densities
\begin{equation}\label{eq:connvardens}
\tilde{Y}_{\mu}{}^{\nu\tau} = Y_{\mu}{}^{\nu\tau}\sqrt{-g}\,, \quad
\tilde{H}_{\mu}{}^{\nu\tau} = H_{\mu}{}^{\nu\tau}\sqrt{-g}\,.
\end{equation}
Expanding the distortion into the nonmetricity and the torsion, and using the fact that the covariant derivative of the density factor with respect to the teleparallel connection is given by
\begin{equation}\label{eq:covderdens}
\nabla_{\mu}\sqrt{-g} = \frac{1}{2}g^{\nu\rho}\nabla_{\mu}g_{\nu\rho}\sqrt{-g} = \frac{1}{2}Q_{\mu\nu}{}^{\nu}\sqrt{-g} = M^{\nu}{}_{\nu\mu}\sqrt{-g}\,,
\end{equation}
one finds that the connection equation can equivalently be written as
\begin{equation}\label{eq:genconnfielddens}
\nabla_{\tau}\tilde{Y}_{\mu}{}^{\nu\tau} - T^{\omega}{}_{\omega\tau}\tilde{Y}_{\mu}{}^{\nu\tau} = \nabla_{\tau}\tilde{H}_{\mu}{}^{\nu\tau} - T^{\omega}{}_{\omega\tau}\tilde{H}_{\mu}{}^{\nu\tau}\,.
\end{equation}
This latter form is particularly convenient if one considers a symmetric teleparallel gravity theory instead of the general teleparallel case. Imposing vanishing torsion, either by introducing another Lagrange multiplier into the action or further restricting the variation, then yields the field equations
\begin{equation}\label{eq:symtelefield}
W_{\mu\nu} = \Theta_{\mu\nu}\,, \quad
\nabla_{\nu}\nabla_{\rho}\tilde{Y}_{\mu}{}^{\nu\rho} = \nabla_{\nu}\nabla_{\rho}\tilde{H}_{\mu}{}^{\nu\rho}\,.
\end{equation}
Finally, we also mention the case of metric teleparallel gravity, where imposing vanishing nonmetricity leads to a single combined field equation given by
\begin{equation}\label{eq:mettelefield}
W^{\mu\nu} - \nabla_{\rho}Y^{\mu\nu\rho} + Y^{\mu\nu\rho}T^{\tau}{}_{\tau\rho} = \Theta^{\mu\nu} - \nabla_{\rho}H^{\mu\nu\rho} + H^{\mu\nu\rho}T^{\tau}{}_{\tau\rho}\,.
\end{equation}
It follows from the structure of the field equations, which in turn follows from the flatness condition on the teleparallel connection, that the variation with respect to the connection enters only via the terms
\begin{equation}\label{eq:redhypmom}
Z_{\mu}{}^{\nu} = W_{\mu}{}^{\nu} - \nabla_{\tau}Y_{\mu}{}^{\nu\tau} + M^{\omega}{}_{\tau\omega}Y_{\mu}{}^{\nu\tau}\,, \quad
I_{\mu}{}^{\nu} = \Theta_{\mu}{}^{\nu} - \nabla_{\tau}H_{\mu}{}^{\nu\tau} + M^{\omega}{}_{\tau\omega}H_{\mu}{}^{\nu\tau}\,,
\end{equation}
or equivalently their respective densities
\begin{equation}
\tilde{Z}_{\mu}{}^{\nu} = \tilde{W}_{\mu}{}^{\nu} - \nabla_{\tau}\tilde{Y}_{\mu}{}^{\nu\tau} + T^{\omega}{}_{\omega\tau}\tilde{Y}_{\mu}{}^{\nu\tau}\,, \quad
\tilde{I}_{\mu}{}^{\nu} = \tilde{\Theta}_{\mu}{}^{\nu} - \nabla_{\tau}\tilde{H}_{\mu}{}^{\nu\tau} + T^{\omega}{}_{\omega\tau}\tilde{H}_{\mu}{}^{\nu\tau}\,,
\end{equation}
where we have also included the metric variation, as it turns out to lead to a simpler linear combination of the field equations, as we will see below. Note in particular that in the commonly considered case of vanishing coupling of the matter to the teleparallel connection, we have vanishing hypermomentum \(H_{\mu}{}^{\nu\rho} = 0\), and thus \(I_{\mu\nu} = \Theta_{\mu\nu}\). In terms of these quantities, the general teleparallel field equations take the form
\begin{equation}\label{eq:gentelefield2}
W_{\mu\nu} = \Theta_{\mu\nu}\,, \quad
Z_{\mu\nu} = I_{\mu\nu}\,,
\end{equation}
the symmetric teleparallel field equations read
\begin{equation}\label{eq:symtelefield2}
W_{\mu\nu} = \Theta_{\mu\nu}\,, \quad
\nabla_{\nu}\tilde{Z}_{\mu}{}^{\nu} = \nabla_{\nu}\tilde{I}_{\mu}{}^{\nu}\,,
\end{equation}
and finally the single metric teleparallel field equation becomes
\begin{equation}\label{eq:mettelefield2}
Z_{\mu\nu} = I_{\mu\nu}\,.
\end{equation}
Note that indices of terms which appear under a covariant derivative can be raised and lowered only in case of a metric compatible connection. The full virtue of writing the teleparallel field equations and their corresponding matter sources for the discussion of the cosmological background and perturbations lies in the fact that by studying their cosmological perturbative expansion of \(W_{\mu\nu}, Z_{\mu\nu}, \Theta_{\mu\nu}, I_{\mu\nu}\), we cover all flavors of teleparallel gravity theories. This will be done in the following sections.

\section{Homogeneous and isotropic background}\label{sec:background}
We now turn our focus to a brief review of the geometric structure of the homogeneous and isotropic background cosmology in teleparallel gravity, and introduce the notation and conventions we use in this article. We start with the metric geometry, which is the well-known Friedmann-Lemaître-Robertson-Walker geometry, in section~\ref{ssec:cosmometric}. We then display the various teleparallel cosmologies in section~\ref{ssec:cosmotele}. The cosmologically symmetric matter sector is discussed in section~\ref{ssec:cosmoenmom}. In section~\ref{ssec:cosmofeq} we then come to the cosmologically symmetric field equations. Finally, in section~\ref{ssec:tensordec} we discuss how to decompose tensors into their time and space components by making use of the homogeneous and isotropic background geometry.

\subsection{Homogeneous and isotropic metric geometry}\label{ssec:cosmometric}
We now briefly review the most general teleparallel geometry which obeys the cosmological symmetry. Using spherical coordinates \((x^{\mu}) = (t, r, \vartheta, \varphi)\), this condition means that all geometric objects introduced above are invariant under the three generators of rotations
\begin{equation}
\sin\varphi\partial_{\vartheta} + \frac{\cos\varphi}{\tan\vartheta}\partial_{\varphi}\,, \quad
-\cos\varphi\partial_{\vartheta} + \frac{\sin\varphi}{\tan\vartheta}\partial_{\varphi}\,, \quad
-\partial_{\varphi}\,,
\end{equation}
as well as the three translation generators
\begin{subequations}
\begin{align}
& \chi\sin\vartheta\cos\varphi\partial_r + \frac{\chi}{r}\cos\vartheta\cos\varphi\partial_{\vartheta} - \frac{\chi\sin\varphi}{r\sin\vartheta}\partial_{\varphi}\,,\\
& \chi\sin\vartheta\sin\varphi\partial_r + \frac{\chi}{r}\cos\vartheta\sin\varphi\partial_{\vartheta} + \frac{\chi\cos\varphi}{r\sin\vartheta}\partial_{\varphi}\,,\\
& \chi\cos\vartheta\partial_r - \frac{\chi}{r}\sin\vartheta\partial_{\vartheta}\,,
\end{align}
\end{subequations}
where \(\chi = \sqrt{1 - r^2u^2}\) and \(u\) is a real or imaginary constant which specifies the spatial curvature of the background metric. Imposing this symmetry on the metric leads to the well-known result that it must be of the Friedmann-Lemaître-Robertson-Walker form, which can be written as
\begin{equation}\label{eq:metricsplit}
\bar{g}_{\mu\nu} = -n_{\mu}n_{\nu} + h_{\mu\nu}\,,
\end{equation}
and we use a bar on top in order to denote background values of tensor fields. Here we have introduced the unit normal covector field
\begin{equation}
n_{\mu}\dd x^{\mu} = -N\dd t
\end{equation}
and induced spatial metric
\begin{equation}
h_{\mu\nu}\dd x^{\mu} \otimes \dd x^{\nu} = A^2\gamma_{ab}\dd x^a \otimes \dd x^b
\end{equation}
on the constant time hypersurfaces, where
\begin{equation}
\gamma_{ab}\dd x^a \otimes \dd x^b = \frac{\dd r \otimes \dd r}{\chi^2} + r^2(\dd\vartheta \otimes \dd\vartheta + \sin^2\vartheta\dd\varphi \otimes \dd\varphi)
\end{equation}
is the maximally symmetric metric on the spatial hypersurfaces corresponding to the choice of the constant \(u\). The metric is thus fully determined by two functions of time, which we call the lapse function \(N = N(t)\) and scale factor \(A = A(t)\), and the curvature parameter \(u\). The background metric \(\bar{g}_{\mu\nu}\) further defines a totally antisymmetric tensor \(\bar{\epsilon}_{\mu\nu\rho\sigma}\), which is normalized such that
\begin{equation}
\bar{\epsilon}_{0123} = \sqrt{-\bar{g}} = \frac{NA^3r^2\sin\vartheta}{\chi}\,.
\end{equation}
We denote by \(\bar{\varepsilon}_{\mu\nu\rho}\) its purely spatial part defined by
\begin{equation}
\bar{\varepsilon}_{\mu\nu\rho} = n^{\sigma}\bar{\epsilon}_{\sigma\mu\nu\rho}\,, \quad
\bar{\epsilon}_{\mu\nu\rho\sigma} = 4\bar{\varepsilon}_{[\mu\nu\rho}n_{\sigma]}\,,
\end{equation}
which can also be expressed as
\begin{equation}
\bar{\varepsilon}_{\mu\nu\rho}\dd x^{\mu} \otimes \dd x^{\nu} \otimes \dd x^{\rho} = A^3\upsilon_{abc}\dd x^a \otimes \dd x^b \otimes \dd x^c
\end{equation}
through the totally antisymmetric tensor \(\upsilon_{abc}\) of the metric \(\gamma_{ab}\). The latter is normalized such that
\begin{equation}
\upsilon_{123} = \sqrt{-\gamma} = \frac{r^2\sin\vartheta}{\chi}\,.
\end{equation}
We will make use of these quantities in order to decompose other tensor fields into temporal and spatial components in the following sections.

\subsection{Homogeneous and isotropic teleparallel backgrounds}\label{ssec:cosmotele}
In order to specify the cosmologically symmetric teleparallel connection, we can make use of the decomposition~\eqref{eq:conndec}, from which we know that the connection is characterized through its torsion and nonmetricity. Imposing the cosmological symmetry, we find that the most general teleparallel background geometry is of the form
\begin{subequations}
\begin{align}
\bar{T}^{\mu}{}_{\nu\rho} &= \frac{2}{A}(\mathcal{T}_1h^{\mu}_{[\nu}n_{\rho]} + \mathcal{T}_2\bar{\varepsilon}^{\mu}{}_{\nu\rho})\,,\\
\bar{Q}_{\rho\mu\nu} &= \frac{2}{A}(\mathcal{Q}_1n_{\rho}n_{\mu}n_{\nu} + 2\mathcal{Q}_2n_{\rho}h_{\mu\nu} + 2\mathcal{Q}_3h_{\rho(\mu}n_{\nu)})\,,
\end{align}
\end{subequations}
and thus determined by five further functions \(\mathcal{T}_1, \mathcal{T}_2, \mathcal{Q}_1, \mathcal{Q}_2, \mathcal{Q}_3\) of time. Note, however, that the five parameter functions introduced above in the torsion and nonmetricity cannot be chosen independently, but must further be restricted by the condition~\eqref{eq:nocurv} of vanishing curvature. It turns out that there are five possible solutions to this condition~\cite{Hohmann:2022mlc,Heisenberg:2022mbo}. For \(u \neq 0\), we find the two branches
\begin{equation}\label{eq:cosmogaxi}
\mathcal{T}_2 = \pm u\,, \quad
\mathcal{T}_1 - \mathcal{Q}_2 = \mathcal{H}\,, \quad
\mathcal{Q}_3 = 0
\end{equation}
and
\begin{equation}\label{eq:cosmogvec}
\mathcal{T}_2 = 0\,, \quad
(\mathcal{H} - \mathcal{T}_1 + \mathcal{Q}_2)(\mathcal{H} - \mathcal{T}_1 + \mathcal{Q}_2 - \mathcal{Q}_3) = -u^2\,, \quad
\mathcal{Q}_1 + \mathcal{Q}_2 = -\frac{\mathcal{H}' - \mathcal{T}_1' + \mathcal{Q}_2'}{\mathcal{H} - \mathcal{T}_1 + \mathcal{Q}_2}\,,
\end{equation}
while for \(u = 0\) there are the three solutions
\begin{equation}\label{eq:cosmogcoin}
\mathcal{T}_2 = 0\,, \quad
\mathcal{T}_1 - \mathcal{Q}_2 = \mathcal{H}\,, \quad
\mathcal{Q}_3 = 0\,,
\end{equation}
\begin{equation}\label{eq:cosmogconf}
\mathcal{T}_2 = 0\,, \quad
\mathcal{T}_1 - \mathcal{Q}_2 = \mathcal{H}\,, \quad
\mathcal{Q}_1 + \mathcal{Q}_2 = \frac{\mathcal{Q}_3'}{\mathcal{Q}_3}\,.
\end{equation}
and
\begin{equation}\label{eq:cosmogpara}
\mathcal{T}_2 = 0\,, \quad
\mathcal{T}_1 - \mathcal{Q}_2 + \mathcal{Q}_3 = \mathcal{H}\,, \quad
\mathcal{Q}_1 + \mathcal{Q}_2 = -\frac{\mathcal{Q}_3'}{\mathcal{Q}_3}
\end{equation}
Here, \(\mathcal{H}\) denotes the conformal Hubble parameter
\begin{equation}
\mathcal{H} = \frac{A'}{A} = \frac{1}{N}\frac{\dd A}{\dd t}\,,
\end{equation}
where a prime denotes the conformal time derivative
\begin{equation}
F' = \frac{A}{N}\frac{\dd F}{\dd t}
\end{equation}
for any function \(F = F(t)\) of time. We finally remark that the special cases of symmetric teleparallel geometries and metric teleparallel geometries can be obtained from the aforementioned five solutions by restriction to either vanishing torsion~\cite{Hohmann:2021ast,DAmbrosio:2021pnd} or vanishing nonmetricity~\cite{Hohmann:2020zre}.

\subsection{Energy-momentum-hypermomentum}\label{ssec:cosmoenmom}
Demanding that the background energy-momentum tensor \(\bar{\Theta}_{\mu\nu}\) satisfies the conditions of homogeneity and isotropy, one finds that it must be of the familiar form
\begin{equation}\label{eq:cosmoenmom}
\bar{\Theta}_{\mu\nu} = \bar{\rho}n_{\mu}n_{\nu} + \bar{p}h_{\mu\nu}\,,
\end{equation}
where \(\bar{\rho}\) and \(\bar{p}\) are the background values of the density and pressure. An analogous consideration can be applied to the (reduced) hypermomentum, which we introduced in section~\ref{ssec:genactfield}, and we find that it must be of the form
\begin{equation}\label{eq:cosmohyp}
\bar{I}_{\mu\nu} = \bar{\varrho}n_{\mu}n_{\nu} + \bar{\pi}h_{\mu\nu}\,,
\end{equation}
with two further background scalars \(\bar{\varrho}\) and \(\bar{\pi}\). By analogy, we may call these quantities hyperdensity and hyperpressure.

\subsection{Teleparallel gravity field equations}\label{ssec:cosmofeq}
In full analogy with the energy-momentum variables discussed above, we can also express the cosmological background value of the terms \(W_{\mu\nu}\) and \(Z_{\mu\nu}\) appearing on the gravitational side of the field equations through the background geometry. These then take the form
\begin{equation}\label{eq:cosmofeq}
\bar{W}_{\mu\nu} = \mathfrak{N}n_{\mu}n_{\nu} + \mathfrak{H}h_{\mu\nu}\,, \quad
\bar{Z}_{\mu\nu} = \mathfrak{T}n_{\mu}n_{\nu} + \mathfrak{S}h_{\mu\nu}\,,
\end{equation}
where we have introduced four functions \(\mathfrak{N}, \mathfrak{H}, \mathfrak{T}, \mathfrak{S}\) of time, which must be derived from the gravitational part of the action for any particular teleparallel gravity theory to be studied. In terms of these, we can express the cosmological background equations as follows. For the general teleparallel case~\eqref{eq:gentelefield2}, we have the equations
\begin{equation}
\mathfrak{N} = \bar{\rho}\,, \quad
\mathfrak{H} = \bar{p}\,, \quad
\mathfrak{T} = \bar{\varrho}\,, \quad
\mathfrak{S} = \bar{\pi}\,.
\end{equation}
For the symmetric teleparallel case~\eqref{eq:symtelefield2}, we find
\begin{equation}
\mathfrak{N} = \bar{\rho}\,, \quad
\mathfrak{H} = \bar{p}\,, \quad
\mathfrak{T}' + 3\mathcal{H}(\mathfrak{T} + \mathfrak{S}) = \bar{\varrho}' + 3\mathcal{H}(\bar{\varrho} + \bar{\pi})\,.
\end{equation}
Finally, for the metric teleparallel case~\eqref{eq:mettelefield2}, we have
\begin{equation}
\mathfrak{T} = \bar{\varrho}\,, \quad
\mathfrak{S} = \bar{\pi}\,.
\end{equation}
One of the fundamental principles of perturbation theory is to solve the field equations by increasing order, starting from the background field equations. When studying linear perturbations, we therefore assume that the background geometry satisfies the background field equations given above. This will be discussed further when we arrive at the perturbed field equations.

\subsection{Tensor decomposition}\label{ssec:tensordec}
In the following sections, we will frequently make use of the decomposition of tensor fields into their temporal and spatial parts as shown in~\cite{Hohmann:2020vcv}. For this purpose, it is convenient to introduce the spatial tensor fields
\begin{equation}
\Pi_{\mu}^a\partial_a \otimes \dd x^{\mu} = A\delta^a_b\,\partial_a \otimes \dd x^b\,, \quad
\Pi^{\mu}_a\partial_{\mu} \otimes \dd x^a = A^{-1}\delta_a^b\,\partial_b \otimes \dd x^a\,.
\end{equation}
One can easily see that they are related to the unit conormal \(n_{\mu}\) and induced spatial metric \(h_{\mu\nu}\) by
\begin{equation}
n_{\mu}\Pi^{\mu}_a = 0\,, \quad
n^{\mu}\Pi_{\mu}^a = 0\,, \quad
h_{\mu\nu}\Pi^{\mu}_a\Pi^{\nu}_b = \gamma_{ab}\,, \quad
\gamma_{ab}\Pi_{\mu}^a\Pi_{\nu}^b = h_{\mu\nu}\,.
\end{equation}
Using the spatial tensor fields and the unit conormal, we can now decompose tensor fields of arbitrary rank. For a vector field \(X = X^{\mu}\partial_{\mu}\) we introduce the notation
\begin{equation}
X = N^{-1}\hat{X}^0\partial_t + A^{-1}\hat{X}^a\partial_a\,, \quad
\hat{X}^0 = -n_{\mu}X^{\mu} = NX^0\,, \quad
\hat{X}^a = \Pi_{\mu}^aX^{\mu} = AX^a
\end{equation}
for the temporal and spatial components. Conversely, for a covector field \(\alpha = \alpha_{\mu}\dd x^{\mu}\) we write
\begin{equation}
\alpha = N\hat{\alpha}_0\,\dd t + A\hat{\alpha}_a\,\dd x^a\,, \quad
\hat{\alpha}_0 = n^{\mu}\alpha_{\mu} = N^{-1}\alpha_0\,, \quad
\hat{\alpha}_a = \Pi^{\mu}_a\alpha_{\mu} = A^{-1}\alpha_a\,.
\end{equation}
These quantities are defined such that when spacetime indices of the tensor fields \(X^{\mu}\) and \(\alpha_{\mu}\) are raised and lowered with the background metric \(\bar{g}_{\mu\nu}\), the temporal and spatial indices of the decomposed fields are raised and lowered with the background metric \(-\dd t \otimes \dd t + \gamma_{ab}\dd x^a \otimes \dd x^b\),
\begin{equation}
\hat{X}_0 = -\hat{X}^0\,, \quad
\hat{X}_a = \gamma_{ab}\hat{X}^b\,, \quad
\hat{\alpha}^0 = -\hat{\alpha}_0\,, \quad
\hat{\alpha}^a = \gamma^{ab}\hat{\alpha}_b\,.
\end{equation}
The advantage of this definition is that the metric used for raising and lowering indices does not depend on the time coordinate, and so raising and lowering of indices commutes with taking time derivatives. Using this decomposition, we can also decompose the covariant derivative of tensor fields. For this purpose, we denote by \(\dd_a\) the Levi-Civita covariant derivative of the spatial metric \(\gamma_{ab}\). One finds that its connection coefficients are given by
\begin{equation}
\frac{1}{2}\gamma^{ad}(\partial_b\gamma_{dc} + \partial_c\gamma_{bd} - \partial_d\gamma_{bc}) = \frac{1}{2}\bar{g}^{a\mu}(\partial_b\bar{g}_{\mu c} + \partial_c\bar{g}_{b\mu} - \partial_{\mu}\bar{g}_{bc}) = \lc{\bar{\Gamma}}^a{}_{bc}\,,
\end{equation}
since the spatial components of the two metric are conformally related by the scale factor which depends only on time, and is thus inert with respect to purely spatial derivatives. Evaluating also the remaining connection coefficients of the Levi-Civita connection, one finds that the covariant derivative of a vector field decomposes as
\begin{equation}
\lc{\bar{\nabla}}_{\mu}X^{\nu} = -N^{-1}n_{\mu}(n^{\nu}\partial_t\hat{X}^0 + \Pi^{\nu}_a\partial_t\hat{X}^a) + A^{-1}\Pi_{\mu}^a(n^{\nu}\dd_a\hat{X}^0 + \Pi^{\nu}_b\dd_a\hat{X}^b) + H(h_{\mu}^{\nu}\hat{X}^0 + \gamma_{ab}\Pi_{\mu}^an^{\nu}\hat{X}^b)\,,
\end{equation}
where the Hubble parameter is given by
\begin{equation}
H = \frac{1}{NA}\frac{\dd A}{\dd t} = \frac{\mathcal{H}}{A}\,.
\end{equation}
Separated into time and space components, one thus has
\begin{equation}
\lc{\bar{\nabla}}_{\mu}X^{\nu}\dd x^{\mu} \otimes \partial_{\nu} = N^{-1}[\partial_t\hat{X}^0\dd t + (\dd_a\hat{X}^0 + \mathcal{H}\hat{X}_a)\dd x^a] \otimes \partial_t + A^{-1}[\partial_t\hat{X}^b\dd t + (\dd_a\hat{X}^b + \mathcal{H}\hat{X}^0\delta_a^b)\dd x^a] \otimes \partial_b\,.
\end{equation}
Similarly, one can decompose the covariant derivative \(\bar{\nabla}_{\mu}\) with respect to the background of the teleparallel connection, by making use of the decomposition~\eqref{eq:conndec}. These formulas are useful to decompose the perturbations of the torsion and nonmetricity tensors and their derivatives, which appear in the perturbed cosmological field equations.

\section{Cosmological perturbations}\label{sec:cosmotelepert}
An important part of the theory of cosmological perturbations is to decompose the perturbations of the fundamental field variables into components which are irreducible under the rotation group. This is done by first performing an algebraic decomposition into temporal and spatial parts, as well as traces and trace-free parts, followed by a differential decomposition into pure divergences and divergence-free tensors. For the space-time decomposition, we make use of the procedure outlined in section~\ref{ssec:tensordec}.

\subsection{Metric}\label{ssec:metpert}
We start with the decomposition of the metric perturbations, which we write in the form
\begin{equation}\label{eq:metricpot}
\hat{\varsigma}_{00} = -2\hat{\phi}\,, \quad
\hat{\varsigma}_{0a} = \dd_a\hat{b} + \hat{s}_a\,, \quad
\hat{\varsigma}_{ab} = -2\psi\gamma_{ab} + 2\dd_a\dd_be + 2\dd_{(a}f_{b)} + \hat{q}_{ab}\,,
\end{equation}
into four scalars \(\hat{\phi}, \hat{\psi}, \hat{b}, \hat{e}\), two divergence-free vectors \(\hat{s}_a, \hat{f}_a\) and one trace-free, divergence-free, symmetric tensor \(\hat{q}_{ab}\). The latter thus satisfy the conditions
\begin{equation}
\dd^a\hat{s}_a = \dd^a\hat{f}_a = 0\,, \quad
\dd^a\hat{q}_{ab} = 0\,, \quad
\hat{q}_{[ab]} = 0\,, \quad
\hat{q}^a{}_a = 0\,.
\end{equation}
Note that this is simply the standard decomposition of the metric perturbations known from cosmological perturbation theory~\cite{}.

\subsection{Connection}\label{ssec:connpert}
Similarly to the decomposition of the metric perturbations, we can decompose the perturbation \(\lambda_{\mu\nu}\) of the teleparallel connection, where we have lowered one index with the background metric \(\bar{g}_{\mu\nu}\) for convenience. Here it is helpful to consider the symmetric and antisymmetric parts independently, and thus to define
\begin{equation}\label{eq:connpot}
\hat{\lambda}_{00} = -\hat{\varphi}\,, \quad
\hat{\lambda}_{a0} + \hat{\lambda}_{0a} = \dd_a\hat{j} + \hat{c}_a\,, \quad
\hat{\lambda}_{a0} - \hat{\lambda}_{0a} = \dd_a\hat{y} + \hat{v}_a\,, \quad
\hat{\lambda}_{ab} = -\varpi\gamma_{ab} + \dd_a\dd_b\sigma + \dd_{(a}z_{b)} + \frac{1}{2}\hat{u}_{ab} + \upsilon_{abc}(\hat{w}^c + \dd^c\xi)\,.
\end{equation}
Here, \(\hat{\varphi}, \hat{\varpi}, \hat{j}, \hat{y}, \hat{\sigma}\) are scalars, \(\hat{\xi}\) is a pseudoscalar, \(\hat{c}_a, \hat{v}_a, \hat{z}_a\) are divergence-free vectors, \(\hat{w}_a\) is a divergence-free pseudovector and \(\hat{u}_{ab}\) is a symmetric, trace-free, divergence-free tensor. The advantage of this choice of the decomposition becomes clear when we consider the cases of metric and symmetric teleparallel gravity. For the former, vanishing nonmetricity leads to the condition \(\varsigma_{\mu\nu} = 2\lambda_{(\mu\nu)}\), which means that we can find the conditions
\begin{equation}
\hat{\varphi} = \hat{\phi}\,, \quad
\hat{\varpi} = \hat{\psi}\,, \quad
\hat{\sigma} = \hat{e}\,, \quad
\hat{j} = \hat{b}\,, \quad
\hat{z}_a = \hat{f}_a\,, \quad
\hat{c}_a = \hat{s}_a\,, \quad
\hat{u}_{ab} = \hat{q}_{ab}
\end{equation}
on the connection perturbations, leaving only \(\hat{y}, \hat{\xi}, \hat{v}_a, \hat{w}_a\) as its independent components, while the remaining ones are expressed through the metric perturbations. Finally, for the symmetric teleparallel case, in which the connection perturbation takes the form \(\lambda^{\mu}{}_{\nu} = \nabla_{\nu}\zeta^{\mu}\), one can decompose \(\zeta^{\mu}\) in the form
\begin{equation}\label{eq:symconnpot}
\hat{\zeta}_0 = A\hat{\alpha}\,, \quad
\hat{\zeta}_a = A(\dd_a\hat{\beta} + \hat{\varkappa}_a)\,,
\end{equation}
where we included another scale factor for convenience, as it will cancel a corresponding factor incurred from the additional derivative. In this case the fundamental perturbation variables are those originating from the metric perturbation, as well as the scalar \(\hat{\alpha}, \hat{\beta}\) and the divergence-free vector \(\hat{\varkappa}_a\).

\subsection{Energy-momentum}\label{ssec:enmompert}
We now come to the perturbation of the energy-momentum tensor \(\Theta_{\mu\nu}\) around its homogeneous and isotropic background~\eqref{eq:cosmoenmom}. To define the irreducible components of the perturbation, we follow the method outlined in~\cite{Kodama:1984ziu,Malik:2008im}, which we show here in detail, in order to generalize it for the hypermomentum in the next section. First note that the background energy-momentum tensor~\eqref{eq:cosmoenmom} satisfies
\begin{equation}
\bar{\Theta}^{\mu}{}_{\nu}n^{\nu} = -\bar{\rho}n^{\mu}\,,
\end{equation}
and so \(n^{\mu}\) is a timelike eigenvector of \(\bar{\Theta}^{\mu}{}_{\nu}n^{\nu}\) with eigenvalue \(-\bar{\rho}\). Further, every vector \(V^{\mu}\) which satisfies \(V^{\mu}n_{\mu} = 0\), and is thus tangent to the constant time spatial hypersurfaces, satisfies
\begin{equation}
\bar{\Theta}^{\mu}{}_{\nu}V^{\nu} = \bar{p}V^{\mu}\,,
\end{equation}
and is thus an eigenvector with eigenvalue \(p\). The trace is
\begin{equation}
\bar{\Theta}^{\mu}{}_{\mu} = 3\bar{p} - \bar{\rho}\,.
\end{equation}
If we demand that the full, perturbed energy-momentum tensor is given by a small perturbation of this background, it retains the property that it has one timelike eigenvector. Denoting this eigenvector by \(u^{\mu}\), and demanding that it is normalized as \(u^{\mu}u^{\nu}g_{\mu\nu} = -1\), its eigenvalue by \(-\rho\) and the trace by \(3p - \rho\), it follows that the energy-momentum tensor can be written as
\begin{equation}
\Theta_{\mu\nu} = (\rho + p)u_{\mu}u_{\nu} + pg_{\mu\nu} + \Pi_{\mu\nu}\,,
\end{equation}
where the last term is the anisotropic stress and satisfies
\begin{equation}\label{eq:anisocond}
u^{\mu}\Pi_{\mu\nu} = 0\,, \quad
\Pi^{\mu}{}_{\mu} = 0\,, \quad
\Pi_{[\mu\nu]} = 0\,.
\end{equation}
Note that \(\rho\) and \(p\) are scalars which are uniquely defined through the eigenvalue of the only timelike eigenvector and the trace of the energy-momentum tensor, and so we can use them to define the perturbations
\begin{equation}
\rho = \bar{\rho} + \hat{\mathcal{E}}\,, \quad
p = \bar{p} + \hat{\mathcal{P}}\,.
\end{equation}
Also the four-velocity is uniquely defined as the unique normalized timelike eigenvector. The normalization condition determines its time component as
\begin{equation}
u^0 = N^{-1}(1 - \phi)\,, \quad
u_0 = -N(1 + \phi)\,.
\end{equation}
Its spatial components can be decomposed into a pure divergence and a divergence-free vector as
\begin{equation}
u^a = A^{-1}(\dd^a\hat{\mathcal{L}} + \hat{\mathcal{X}}^a)\,, \quad
u_a = A(\dd_a\hat{\mathcal{L}} + \hat{\mathcal{X}}_a + \dd_a\hat{b} + \hat{s}_a)\,.
\end{equation}
Finally, the anisotropic stress \(\Pi_{\mu\nu}\) vanishes for the background energy-momentum tensor, and so it is already of linear perturbation order. We can thus replace \(u^{\mu}\) by \(n^{\mu}\) in the condition~\eqref{eq:anisocond}, and conclude that
\begin{equation}
\Pi_{00} = 0\,, \quad
\Pi_{a0} = \Pi_{0a} = 0\,, \quad
\Pi_{[ab]} = 0\,.
\end{equation}
Further, since \(\Pi_{\mu\nu}\) is trace-free, it follows that its spatial components can be decomposed as
\begin{equation}
\Pi_{ab} = A^2\left(\dd_a\dd_b\hat{\mathcal{S}} - \frac{1}{3}\triangle\hat{\mathcal{S}}\gamma_{ab} + \dd_{(a}\hat{\mathcal{V}}_{b)} + \hat{\mathcal{T}}_{ab}\right)\,,
\end{equation}
where \(\hat{\mathcal{S}}\) is a scalar, \(\hat{\mathcal{V}}_a\) is a divergence-free vector and \(\hat{\mathcal{T}}_{ab}\) is a symmetric, divergence-free, trace-free tensor. Defining
\begin{equation}\label{eq:enmompertdef}
\delta\Theta_{\mu\nu} = \Theta_{\mu\nu} - \bar{\Theta}_{\mu\nu} = \theta_{\mu\nu}\,,
\end{equation}
we have thus decomposed the perturbation \(\theta_{\mu\nu}\) into the irreducible components
\begin{subequations}
\begin{align}
\hat{\theta}_{00} &= \hat{\mathcal{E}} - \bar{\rho}\hat{\varsigma}_{00}\,,\\
\hat{\theta}_{0a} &= -\bar{\rho}\hat{\varsigma}_{0a} - (\bar{\rho} + \bar{p})(\dd_a\hat{\mathcal{L}} + \hat{\mathcal{X}}_a)\,,\\
\hat{\theta}_{ab} &= \bar{p}\hat{\varsigma}_{ab} + \hat{\mathcal{P}}\gamma_{ab} + \dd_a\dd_b\hat{\mathcal{S}} - \frac{1}{3}\triangle\hat{\mathcal{S}}\gamma_{ab} + \dd_{(a}\hat{\mathcal{V}}_{b)} + \hat{\mathcal{T}}_{ab}\,,
\end{align}
\end{subequations}
where the metric perturbations are further decomposed as shown in section~\ref{ssec:metpert}.

\subsection{Hypermomentum}\label{ssec:hyppert}
In order to decompose the perturbation of the (reduced) hypermomentum tensor around the cosmologically symmetric background~\eqref{eq:cosmohyp}, we proceed in analogy to the perturbation of the energy-momentum tensor. In contrast to the former, however, the reduced hypermomentum is not symmetric in general, and so its left and right eigenvectors will in general be different. For the background value~\eqref{eq:cosmohyp}, it follows from the homogeneity and isotropy that it must be symmetric, and that \(n^{\mu}\) is both a left and right timelike eigenvector with
\begin{equation}
\bar{I}^{\mu}{}_{\nu}n^{\nu} = n^{\nu}\bar{I}_{\nu}{}^{\mu} = -\bar{\varrho}n^{\mu}\,,
\end{equation}
while the remaining eigenvectors are spacelike. Further, its trace is given by
\begin{equation}
\bar{I}^{\mu}{}_{\mu} = 3\bar{\pi} - \bar{\varrho}\,.
\end{equation}
For the general, perturbed reduced hypermomentum we can thus make the ansatz
\begin{equation}
I_{\mu\nu} = -(\varrho + \pi)\frac{v_{\mu}w_{\nu}}{v_{\rho}w^{\rho}} + \pi g_{\mu\nu} + \Sigma_{\mu\nu}\,,
\end{equation}
with
\begin{equation}\label{eq:anisocondh}
\Sigma_{\mu\nu}v^{\nu} = 0\,, \quad
w^{\mu}\Sigma_{\mu\nu} = 0\,, \quad
\Sigma^{\mu}{}_{\mu} = 0\,,
\end{equation}
and both \(v^{\mu}\) and \(w^{\mu}\) are normalized,
\begin{equation}
v^{\mu}v^{\nu}g_{\mu\nu} = w^{\mu}w^{\nu}g_{\mu\nu} = -1\,.
\end{equation}
If we demand that \(I_{\mu\nu}\) is a small perturbation of \(\bar{I}_{\mu\nu}\), it retains one timelike and three spacelike eigenvectors, although left and right eigenvectors will now differ in general. One finds that
\begin{equation}
I^{\mu}{}_{\nu}v^{\nu} = -\varrho v^{\mu}\,, \quad
w^{\nu}I_{\nu}{}^{\mu} = -\varrho w^{\mu}\,,
\end{equation}
and so the timelike eigenvectors are given by \(v^{\mu}\) and \(w^{\mu}\), with the common eigenvalue \(-\varrho\), while \(\pi\) is defined via the trace
\begin{equation}
I^{\mu}{}_{\mu} = 3\pi - \varrho\,.
\end{equation}
For these two scalars we can thus define the linear perturbations
\begin{equation}
\varrho = \bar{\varrho} + \hat{\mathcal{D}}\,, \quad
\pi = \bar{\pi} + \hat{\mathcal{Q}}\,.
\end{equation}
We then continue with the two timelike eigenvectors. The normalization condition determines their time components
\begin{equation}
v^0 = w^0 = N^{-1}(1 - \phi)\,,
\end{equation}
while the space components define two independent perturbations
\begin{equation}
v^a = A^{-1}(\dd^a\hat{\mathcal{M}} + \hat{\mathcal{Y}}^a)\,, \quad
w^a = A^{-1}(\dd^a\hat{\mathcal{N}} + \hat{\mathcal{Z}}^a)\,.
\end{equation}
Finally, we take a closer look at the anisotropic contribution \(\Sigma_{\mu\nu}\), which is already of linear perturbation order. Hence, the conditions~\eqref{eq:anisocondh} imply
\begin{equation}
\Sigma_{00} = 0\,, \quad
\Sigma_{a0} = \Sigma_{0a} = 0\,.
\end{equation}
We can thus expand it in the form
\begin{equation}
\Sigma_{ab} = A^2\left[\dd_a\dd_b\hat{\mathcal{A}} - \frac{1}{3}\triangle\hat{\mathcal{A}}\gamma_{ab} + \dd_{(a}\hat{\mathcal{B}}_{b)} + \hat{\mathcal{C}}_{ab} + \upsilon_{abc}(\dd^c\hat{\mathcal{I}} + \hat{\mathcal{J}}^c)\right]\,,
\end{equation}
where we now also have an antisymmetric part. In summary, the hypermomentum perturbation splits into the scalars \(\hat{\mathcal{D}}, \hat{\mathcal{Q}}, \hat{\mathcal{M}}, \hat{\mathcal{N}}, \hat{\mathcal{A}}\), the pseudoscalar \(\hat{\mathcal{I}}\), the divergence-free vectors \(\hat{\mathcal{Y}}_a, \hat{\mathcal{Z}}_a, \hat{\mathcal{B}}_a\), the divergence-free pseudovector \(\hat{\mathcal{J}}_a\) and the trace-free, divergence-free, symmetric tensor \(\hat{\mathcal{C}}_{ab}\). Introducing the perturbation
\begin{equation}
\delta I_{\mu\nu} = I_{\mu\nu} - \bar{I}_{\mu\nu} = \iota_{\mu\nu}\,,
\end{equation}
we thus have the expansion
\begin{subequations}
\begin{align}
\hat{\iota}_{00} &= \hat{\mathcal{D}} - \bar{\varrho}\hat{\varsigma}_{00}\,,\\
\hat{\iota}_{a0} &= -\bar{\varrho}\hat{\varsigma}_{a0} - (\bar{\varrho} + \bar{\pi})(\dd_a\hat{\mathcal{M}} + \hat{\mathcal{Y}}_a)\,,\\
\hat{\iota}_{0a} &= -\bar{\varrho}\hat{\varsigma}_{0a} - (\bar{\varrho} + \bar{\pi})(\dd_a\hat{\mathcal{N}} + \hat{\mathcal{Z}}_a)\,,\\
\hat{\iota}_{ab} &= \bar{\pi}\hat{\varsigma}_{ab} + \hat{\mathcal{Q}}\gamma_{ab} + \dd_a\dd_b\hat{\mathcal{A}} - \frac{1}{3}\triangle\hat{\mathcal{A}}\gamma_{ab} + \dd_{(a}\hat{\mathcal{B}}_{b)} + \hat{\mathcal{C}}_{ab} + \upsilon_{abc}(\dd^c\hat{\mathcal{I}} + \hat{\mathcal{J}}^c)\,,
\end{align}
\end{subequations}
where also in this case the metric perturbations are further decomposed as shown in section~\ref{ssec:metpert}.

\subsection{Field equations}\label{ssec:feqpert}
Finally, we also need to provide the perturbative expansion of the terms \(W_{\mu\nu}\) and \(Z_{\mu\nu}\) obtained from the variation of the gravitational part of the action. Note that these two terms exactly match the matter terms \(\Theta_{\mu\nu}\) and \(I_{\mu\nu}\) in the field equations, which we have also used in section~\ref{ssec:cosmofeq} for the cosmologically symmetric background field equations. Hence, it turns out to be convenient to also exactly match their linear perturbative expansion. Defining the perturbations as
\begin{equation}
\delta W_{\mu\nu} = W_{\mu\nu} - \bar{W}_{\mu\nu} = \mathfrak{W}_{\mu\nu}\,, \quad
\delta Z_{\mu\nu} = Z_{\mu\nu} - \bar{Z}_{\mu\nu} = \mathfrak{Z}_{\mu\nu}\,,
\end{equation}
we thus write their irreducible decomposition in the form
\begin{subequations}
\begin{align}
\hat{\mathfrak{W}}_{00} &= \hat{\mathfrak{e}} - \mathfrak{N}\hat{\varsigma}_{00}\,,\\
\hat{\mathfrak{W}}_{0a} &= -\mathfrak{N}\hat{\varsigma}_{0a} - (\mathfrak{N} + \mathfrak{H})(\dd_a\hat{\mathfrak{l}} + \hat{\mathfrak{x}}_a)\,,\\
\hat{\mathfrak{W}}_{ab} &= \mathfrak{H}\hat{\varsigma}_{ab} + \hat{\mathfrak{p}}\gamma_{ab} + \dd_a\dd_b\hat{\mathfrak{s}} - \frac{1}{3}\triangle\hat{\mathfrak{s}}\gamma_{ab} + \dd_{(a}\hat{\mathfrak{v}}_{b)} + \hat{\mathfrak{t}}_{ab}\,,
\end{align}
\end{subequations}
and
\begin{subequations}
\begin{align}
\hat{\mathfrak{Z}}_{00} &= \hat{\mathfrak{d}} - \mathfrak{T}\hat{\varsigma}_{00}\,,\\
\hat{\mathfrak{Z}}_{a0} &= -\mathfrak{T}\hat{\varsigma}_{a0} - (\mathfrak{T} + \mathfrak{S})(\dd_a\hat{\mathfrak{m}} + \hat{\mathfrak{y}}_a)\,,\\
\hat{\mathfrak{Z}}_{0a} &= -\mathfrak{T}\hat{\varsigma}_{0a} - (\mathfrak{T} + \mathfrak{S})(\dd_a\hat{\mathfrak{n}} + \hat{\mathfrak{z}}_a)\,,\\
\hat{\mathfrak{Z}}_{ab} &= \mathfrak{S}\hat{\varsigma}_{ab} + \hat{\mathfrak{q}}\gamma_{ab} + \dd_a\dd_b\hat{\mathfrak{a}} - \frac{1}{3}\triangle\hat{\mathfrak{a}}\gamma_{ab} + \dd_{(a}\hat{\mathfrak{b}}_{b)} + \hat{\mathfrak{c}}_{ab} + \upsilon_{abc}(\dd^c\hat{\mathfrak{i}} + \hat{\mathfrak{j}}^c)\,,
\end{align}
\end{subequations}
Assuming that the background field equations are already satisfied by the background geometry and matter variables, it is then straightforward to write down the linear perturbed field equations.

\section{Gauge transformations and gauge-invariant quantities}\label{sec:gaugetrans}
We now study the transformation of the perturbations under infinitesimal coordinate changes of the form
\begin{equation}\label{eq:coordtrans}
x'^{\mu} = x^{\mu} + V^{\mu}(x)\,,
\end{equation}
where the components of the vector field \(V^{\mu}\) are assumed to be sufficiently small that the metric and teleparallel connection retain the character as small perturbations around a fixed cosmological background, which is given by the same expression in the new coordinates. In order to decompose these relations into irreducible components, and thus obtain the transformation of the irreducible perturbation components, we decompose the transformation vector field as
\begin{equation}
\hat{V}_0 = A\hat{X}\,, \quad
\hat{V}_a = A(\dd_a\hat{Y} + \hat{Z}_a)\,,
\end{equation}
where we once again introduced the scale factor for convenience.

\subsection{Metric}\label{ssec:metgatra}
Under this coordinate change, the metric perturbation undergoes the transformation
\begin{equation}\label{eq:mettrans}
\delta_V\varsigma_{\mu\nu} = \varsigma_{\mu\nu} - \varsigma'_{\mu\nu} = (\mathcal{L}_V\bar{g})_{\mu\nu} = 2\lc{\bar{\nabla}}_{(\mu}V_{\nu)}\,,
\end{equation}
Using this decomposition, one finds that the transformation~\eqref{eq:mettrans} of the metric perturbation decomposes as
\begin{subequations}
\begin{align}
\delta_V\hat{\varsigma}_{00} &= 2(\mathcal{H}\hat{X} + \hat{X}')\,,\\
\delta_V\hat{\varsigma}_{0a} &= \dd_a\hat{X} + (\dd_a\hat{Y} + \hat{Z}_a)'\,,\\
\delta_V\hat{\varsigma}_{ab} &= 2(\dd_a\dd_b\hat{Y} + \dd_{(a}\hat{Z}_{b)} - \mathcal{H}\hat{X}\gamma_{ab})\,.
\end{align}
\end{subequations}
By comparison with the decomposition~\eqref{eq:metricpot}, we find that the irreducible components of the perturbation transform as
\begin{equation}
\delta_V\hat{\phi} = -\mathcal{H}\hat{X} - \hat{X}'\,, \quad
\delta_V\hat{b} = \hat{X} + \hat{Y}'\,, \quad
\delta_V\hat{e} = \hat{Y}\,, \quad
\delta_V\hat{\psi} = \mathcal{H}\hat{X}\,, \quad
\delta_V\hat{s}_a = \hat{Z}'_a\,, \quad
\delta_V\hat{f}_a = \hat{Z}_a\,, \quad
\delta_V\hat{q}_{ab} = 0\,.
\end{equation}
Note that only the tensor perturbation \(\hat{q}_{ab}\) is already gauge-invariant, while the remaining components transform non-trivially under gauge transformations. Nevertheless, one can already see from the structure of these transformations that it is possible to find linear combinations of the metric perturbation components which are gauge-invariant; this is the classical approach used in gauge-invariant cosmological perturbation theory, which straightforwardly leads to the definition of the gauge-invariant variables in the spatially flat gauge. Here we decide to consider a different, more general approach, and so we defer the discussion of gauge-invariant variables to section~\ref{ssec:gaugeinv}.

\subsection{Connection}\label{ssec:conngatra}
while the connection perturbation transforms as
\begin{equation}
\delta_V\delta\Gamma^{\mu}{}_{\nu\rho} = \delta\Gamma^{\mu}{}_{\nu\rho} - \delta\Gamma'^{\mu}{}_{\nu\rho} = (\mathcal{L}_V\bar{\Gamma})^{\mu}{}_{\nu\rho}\,.
\end{equation}
Here the Lie derivative of the connection coefficients is given by
\begin{equation}
(\mathcal{L}_V\bar{\Gamma})^{\mu}{}_{\nu\rho} = V^{\sigma}\partial_{\sigma}\bar{\Gamma}^{\mu}{}_{\nu\rho} - \partial_{\sigma}V^{\mu}\bar{\Gamma}^{\sigma}{}_{\nu\rho} + \partial_{\nu}V^{\sigma}\bar{\Gamma}^{\mu}{}_{\sigma\rho} + \partial_{\rho}V^{\sigma}\bar{\Gamma}^{\mu}{}_{\nu\sigma} + \partial_{\nu}\partial_{\rho}V^{\mu}
= \bar{\nabla}_{\rho}\bar{\nabla}_{\nu}V^{\mu} - V^{\sigma}\bar{R}^{\mu}{}_{\nu\rho\sigma} - \bar{\nabla}_{\rho}(V^{\sigma}\bar{T}^{\mu}{}_{\nu\sigma})\,,
\end{equation}
where the last expression shows that the Lie derivative is a tensor fields. Using the fact that the curvature vanishes and that the teleparallel connection perturbation takes the form~\eqref{eq:connpert}, we find the transformation
\begin{equation}\label{eq:conntrans}
\delta_V\lambda^{\mu}{}_{\nu} = \bar{\nabla}_{\nu}V^{\mu} - V^{\rho}\bar{T}^{\mu}{}_{\nu\rho}
\end{equation}
of the fundamental perturbation variable. Two special cases are worth noting. In the case of vanishing nonmetricity, we can lower one index under the covariant derivative and write the gauge transformation as
\begin{equation}
\delta_V\lambda_{\mu\nu} = \bar{\nabla}_{\nu}V_{\mu} - V^{\rho}\bar{T}_{\mu\nu\rho} = \lc{\bar{\nabla}}_{\nu}V_{\mu} + V^{\rho}\bar{K}_{\mu\nu\rho}\,.
\end{equation}
Using the fact that the contortion is antisymmetric in its first two indices, we find that the symmetric part of this expression is given by
\begin{equation}
\delta_V\lambda_{(\mu\nu)} = \lc{\bar{\nabla}}_{(\mu}V_{\nu)} = \frac{1}{2}\delta_V\varsigma_{\mu\nu}\,,
\end{equation}
which is consistent with the relation~\eqref{eq:connpertm} in this case. In the second special case of vanishing torsion, the relation~\eqref{eq:connperts} implies
\begin{equation}\label{eq:symconntrans}
\delta_V\zeta^{\mu} = V^{\mu}\,.
\end{equation}
We then continue with the gauge transformation of the connection perturbation. Lowering one index and decomposing into space and time components, the transformation~\eqref{eq:conntrans} becomes
\begin{subequations}
\begin{align}
\delta_V\hat{\lambda}_{00} &= (\mathcal{H} - \mathcal{Q}_1)\hat{X} + \hat{X}'\,,\\
\delta_V\hat{\lambda}_{a0} &= (\mathcal{H} + \mathcal{Q}_2 - \mathcal{T}_1)(\dd_a\hat{Y} + \hat{Z}_a) + (\dd_a\hat{Y} + \hat{Z}_a)'\,,\\
\delta_V\hat{\lambda}_{0b} &= \dd_b\hat{X} - (\mathcal{H} - \mathcal{Q}_3 + \mathcal{Q}_2 - \mathcal{T}_1)(\dd_b\hat{Y} + \hat{Z}_b)\,,\\
\delta_V\hat{\lambda}_{ab} &= \dd_a\dd_b\hat{Y} + \dd_b\hat{Z}_a - (\mathcal{H} + \mathcal{Q}_2)\hat{X}\gamma_{ab} - \mathcal{T}_2\upsilon_{abc}(\dd^c\hat{Y} + \hat{Z}^c)\,.
\end{align}
\end{subequations}
Before we can read of the transformation of the irreducible components, we calculate the symmetric and antisymmetric parts, which read
\begin{subequations}
\begin{align}
\delta_V\hat{\lambda}_{a0} + \delta_V\hat{\lambda}_{0a} &= \dd_a\hat{X} + \mathcal{Q}_3(\dd_a\hat{Y} + \hat{Z}_a) + (\dd_a\hat{Y} + \hat{Z}_a)'\,,\\
\delta_V\hat{\lambda}_{a0} - \delta_V\hat{\lambda}_{0a} &= -\dd_a\hat{X} + (2\mathcal{H} - \mathcal{Q}_3 + 2\mathcal{Q}_2 - 2\mathcal{T}_1)(\dd_a\hat{Y} + \hat{Z}_a) + (\dd_a\hat{Y} + \hat{Z}_a)'\,,\\
\delta_V\hat{\lambda}_{(ab)} &= \dd_a\dd_b\hat{Y} + \dd_{(a}\hat{Z}_{b)} - (\mathcal{H} + \mathcal{Q}_2)\hat{X}\gamma_{ab}\,,\\
\delta_V\hat{\lambda}_{[ab]} &= -\dd_{[a}\hat{Z}_{b]} - \mathcal{T}_2\upsilon_{abc}(\dd^c\hat{Y} + \hat{Z}^c)\,.
\end{align}
\end{subequations}
Finally, we need to suitably transform the first term in the last line. Here we use the property of the totally antisymmetric tensor that its contraction with itself gives a generalized (antisymmetric) Kronecker symbol, and so we can write
\begin{equation}
\dd_{[a}\hat{Z}_{b]} = \frac{1}{2}\upsilon_{abc}\upsilon^{dec}\dd_d\hat{Z}_e\,.
\end{equation}
Now the expression \(\upsilon^{dec}\dd_d\hat{Z}_e\) is a pseudovector. We then make use of the first Bianchi identity to calculate
\begin{equation}
\dd_c(\upsilon^{dec}\dd_d\hat{Z}_e) = \upsilon^{dec}\dd_{[c}\dd_{d]}\hat{Z}_e = \frac{1}{2}\upsilon^{dec}R^f{}_{ecd}\hat{Z}_f = 0\,,
\end{equation}
and so we see that this pseudovector is divergence-free. Hence, it contributes to the transformation of \(\hat{w}_a\) only. In summary, we thus find the transformation of the irreducible components
\begin{gather}
\delta_V\hat{\varphi} = (\mathcal{Q}_1 - \mathcal{H})\hat{X} - \hat{X}'\,, \quad
\delta_V\hat{\varpi} = (\mathcal{Q}_2 + \mathcal{H})\hat{X}\,, \quad
\delta_V\hat{j} = \hat{X} + \mathcal{Q}_3\hat{Y} + \hat{Y}'\,,\nonumber\\
\delta_V\hat{y} = -\hat{X} + (2\mathcal{H} - \mathcal{Q}_3 + 2\mathcal{Q}_2 - 2\mathcal{T}_1)\hat{Y} + \hat{Y}'\,, \quad
\delta_V\hat{\sigma} = \hat{Y}\,, \quad
\delta_V\hat{\xi} = -\mathcal{T}_2\hat{Y}\,, \quad
\delta_V\hat{u}_{ab} = 0\,,\\
\delta_V\hat{c}_a = \mathcal{Q}_3\hat{Z}_a + \hat{Z}_a'\,, \quad
\delta_V\hat{v}_a = (2\mathcal{H} - \mathcal{Q}_3 + 2\mathcal{Q}_2 - 2\mathcal{T}_1)\hat{Z}_a + \hat{Z}_a'\,, \quad
\delta_V\hat{z}_a = \hat{Z}_a\,, \quad
\delta_V\hat{w}_a = -\mathcal{T}_2\hat{Z}_a - \frac{1}{2}\upsilon_{abc}\dd^b\hat{Z}^c\,.\nonumber
\end{gather}
Finally, using the transformation~\eqref{eq:symconntrans} and the decomposition~\eqref{eq:symconnpot} of the symmetric teleparallel perturbation, we find that its irreducible components transform as
\begin{equation}
\delta_V\hat{\alpha} = \hat{X}\,, \quad
\delta_V\hat{\beta} = \hat{Y}\,, \quad
\delta_V\hat{\varkappa}_a = \hat{Z}_a\,.
\end{equation}
Hence, we have found the transformation of all irreducible perturbation components under infinitesimal coordinate transformations.

\subsection{Energy-momentum and hypermomentum}\label{ssec:enmomgatra}
We continue with the gauge transformation of the energy-momentum tensor. By explicit calculation it follows that the only non-trivial transformation of the irreducible components is given by
\begin{equation}
\delta_V\hat{\mathcal{E}} = -\hat{X}\bar{\rho}'\,, \quad
\delta_V\hat{\mathcal{P}} = -\hat{X}\bar{p}'\,, \quad
\delta_V\hat{\mathcal{L}} = -\hat{Y}'\,, \quad
\delta_V\hat{\mathcal{X}}_a = -\hat{Z}_a'\,,
\end{equation}
while the components of the anisotropic stress are gauge-invariant,
\begin{equation}
\delta_V\hat{\mathcal{S}} = 0\,, \quad
\delta_V\hat{\mathcal{V}}_a = 0\,, \quad
\delta_V\hat{\mathcal{T}}_{ab} = 0\,.
\end{equation}
Analogously, one finds the transformation of the hypermomentum perturbations given by
\begin{equation}
\delta_V\hat{\mathcal{D}} = -\hat{X}\bar{\varrho}'\,, \quad
\delta_V\hat{\mathcal{Q}} = -\hat{X}\bar{\pi}'\,, \quad
\delta_V\hat{\mathcal{M}} = \delta_V\hat{\mathcal{N}} = -\hat{Y}'\,, \quad
\delta_V\hat{\mathcal{Y}}_a = \delta_V\hat{\mathcal{Z}}_a = -\hat{Z}_a'\,,
\end{equation}
together with the gauge-invariant perturbation components
\begin{equation}
\delta_V\hat{\mathcal{A}} = \delta_V\hat{\mathcal{I}} = 0\,, \quad
\delta_V\hat{\mathcal{B}}_a = \delta_V\hat{\mathcal{J}}_a = 0\,, \quad
\delta_V\hat{\mathcal{C}}_{ab} = 0\,.
\end{equation}
Also for the matter variables, it is possible to perform a transition to gauge-invariant variables, in analogy to the geometry perturbations, as we will see below.

\subsection{Field equations}\label{ssec:feqgatra}
Since we have defined the irreducible components of the field equation variation in full analogy to the perturbation of the matter variables, it follows immediately that also their behavior under gauge transformations is identical. Hence, we find that the perturbation \(\mathfrak{W}_{\mu\nu}\) of the metric variation term \(W_{\mu\nu}\) undergoes the non-trivial gauge transformation
\begin{equation}
\delta_V\hat{\mathfrak{e}} = -\hat{X}\mathfrak{N}'\,, \quad
\delta_V\hat{\mathfrak{p}} = -\hat{X}\mathfrak{H}'\,, \quad
\delta_V\hat{\mathfrak{l}} = -\hat{Y}'\,, \quad
\delta_V\hat{\mathfrak{x}}_a = -\hat{Z}_a'\,,
\end{equation}
while the remaining components transform as
\begin{equation}
\delta_V\hat{\mathfrak{s}} = 0\,, \quad
\delta_V\hat{\mathfrak{v}}_a = 0\,, \quad
\delta_V\hat{\mathfrak{t}}_{ab} = 0\,,
\end{equation}
and are thus gauge-invariant. Similarly, the components of the perturbation \(\mathfrak{Z}_{\mu\nu}\) of the connection variation term \(Z_{\mu\nu}\) obey the transformation
\begin{equation}
\delta_V\hat{\mathfrak{d}} = -\hat{X}\mathfrak{T}'\,, \quad
\delta_V\hat{\mathfrak{q}} = -\hat{X}\mathfrak{S}'\,, \quad
\delta_V\hat{\mathfrak{m}} = \delta_V\hat{\mathfrak{n}} = -\hat{Y}'\,, \quad
\delta_V\hat{\mathfrak{y}}_a = \delta_V\hat{\mathfrak{z}}_a = -\hat{Z}_a'\,,
\end{equation}
together with the gauge-invariant components
\begin{equation}
\delta_V\hat{\mathfrak{a}} = \delta_V\hat{\mathfrak{i}} = 0\,, \quad
\delta_V\hat{\mathfrak{b}}_a = \delta_V\hat{\mathfrak{j}}_a = 0\,, \quad
\delta_V\hat{\mathfrak{c}}_{ab} = 0\,.
\end{equation}
Note that if one imposes the background field equations displayed in section~\ref{ssec:cosmofeq}, then both sides of the perturbative field equations shown in section~\ref{ssec:feqpert} are subject to identical gauge transformations. While each side of the field equations undergoes a non-trivial change due to the gauge transformation, the difference of both sides is a gauge-invariant quantity, i.e., it is the same in all gauges. Gauge-invariance of the field equations thus means that this gauge-invariant quantity must vanish if the field equations are imposed.

\subsection{Gauge-invariant perturbations}\label{ssec:gaugeinv}
In the preceding section we have seen that the irreducible components of the metric and connection perturbations transform non-trivially under infinitesimal coordinate transformations, and so their values depend on the choice of the coordinate system. In order to obtain physical quantities, whose values are independent of the choice of the coordinate system, there are two conceptually different, but mathematically and physically equivalent methods. One approach is to perform \emph{gauge fixing}, i.e., to choose a fixed reference coordinate system, which is determined by imposing the condition that a chosen set of irreducible components takes particular values, and regarding the values of the remaining components as physical quantities. Alternatively, one may construct \emph{gauge-invariant} variables by choosing linear combinations of the irreducible components, such that their gauge transformations mutually cancel. To see that these two approaches are mutually equivalent, note that we can write any of the irreducible components, which we symbolically denote \(\hat{Q}\) here, as a sum
\begin{equation}
\hat{Q} = \hat{\boldsymbol{Q}} - \delta_V\hat{Q}\,,
\end{equation}
where \(\hat{\boldsymbol{Q}}\) is the value of \(\hat{Q}\) in a particular, fixed coordinate system, and \(V\) is the vector field which generates the infinitesimal coordinate transformation from this fixed coordinate system to the one in which \(\hat{Q}\) is expressed. Under an infinitesimal coordinate transformation to a different coordinate system, which is specified by \(V'\) instead of \(V\), only the second term on the right hand side changes, while the first one is invariant. From this point of view, \emph{gauge fixing} means to impose conditions on the values \(\hat{\boldsymbol{Q}}_i\) of a chosen set \(\hat{Q}_i\) of quantities in the fixed coordinate system, such as demanding them to vanish. These conditions, which then read
\begin{equation}
0 \equiv \hat{\boldsymbol{Q}}_i = \hat{Q}_i + \delta_V\hat{Q}_i\,,
\end{equation}
then allow us to solve for the components of the vector field \(V\) from the relation \(\delta_V\hat{Q}_i = \hat{Q}_i\). Hence, knowing the values \(\hat{Q}_i\) in some arbitrary coordinate system, we can express \(V\) in terms of these values, and hence find the transformation from the fixed to the arbitrary coordinate system. We can then apply the inverse of the same transformation to any other quantity \(\hat{Q}_k\), and thus calculate its value
\begin{equation}
\hat{\boldsymbol{Q}}_k = \hat{Q}_k + \delta_V\hat{Q}_k
\end{equation}
in the fixed coordinate system, where the second term on the right hand side is a linear combination of the components of \(V\), and thus expressed through the values \(\hat{Q}_i\). This equation then expresses the left hand side \(\hat{\boldsymbol{Q}}_k\) as a \emph{gauge-invariant} linear combination of \(\hat{Q}_k\) and \(\hat{Q}_i\).

To illustrate the construction given above, we provide a few practical examples. In the following, we will denote a particular gauge choice by a letter under the quantities which are expressed in this gauge. The first example, which is common in cosmological perturbation theory, is the \emph{Newtonian gauge}. It is defined by the conditions
\begin{equation}
\ginv{b}{N} = \ginv{e}{N} = 0\,, \quad \ginv{s}{N}_a = 0\,.
\end{equation}
The components of the vector field \(V\) can then be obtained from
\begin{subequations}
\begin{align}
0 &= \hat{b} + \delta_{\gdep{V}{N}}\hat{b} = \hat{b} + \gdep{X}{N} + \gdep{Y}{N}'\,,\\
0 &= \hat{e} + \delta_{\gdep{V}{N}}\hat{e} = \hat{e} + \gdep{Y}{N}\,,\\
0 &= \hat{s}_a + \delta_{\gdep{V}{N}}\hat{s}_a = \hat{s}_a + \gdep{Z}{N}_a'\,,
\end{align}
\end{subequations}
from which one finds
\begin{equation}
\gdep{X}{N} = \hat{e}' - \hat{b}\,, \quad
\gdep{Y}{N} = -\hat{e}\,, \quad
\gdep{Z}{N}_a' = -\hat{s}_a\,.
\end{equation}
Note that \(\gdep{Z}{N}_a\) is determined only up to an integration constant, which determines its value at a fixed constant-time hypersurface. Another possible choice is the \emph{spatially flat gauge}, in which the spatial part of the metric perturbation contains only the tensor part \(\hat{q}_{ab}\), and so the scalar and vector parts vanish,
\begin{equation}
\ginv{\psi}{F} = \ginv{e}{F} = 0\,, \quad \ginv{f}{F}_a = 0\,.
\end{equation}
Solving for the gauge-transforming vector field, one thus finds
\begin{equation}
\gdep{X}{F} = -\mathcal{H}^{-1}\hat{\psi}\,, \quad
\gdep{Y}{F} = -\hat{e}\,, \quad
\gdep{Z}{F}_a = -\hat{f}_a\,.
\end{equation}
The \emph{synchronous gauge} is defined by the conditions
\begin{equation}
\ginv{\phi}{S} = \ginv{b}{S} = 0\,, \quad \ginv{s}{S}_a = 0\,,
\end{equation}
which yields
\begin{equation}
\gdep{X}{S}' + \mathcal{H}\gdep{X}{S} = \hat{\phi}\,, \quad
\gdep{Y}{S}' = -\hat{b} - \gdep{X}{S}\,, \quad
\gdep{Z}{S}_a' = -\hat{s}_a\,.
\end{equation}
The aforementioned gauges are defined in terms of the metric perturbations only. Another class of gauges can be defined by imposing conditions on the matter perturbations. The \emph{comoving gauge} is defined by
\begin{equation}
\ginv{b}{C} = \ginv{\mathcal{L}}{C} = 0\,, \quad \ginv{\mathcal{X}}{C}_a = 0\,,
\end{equation}
which leads to the transformation
\begin{equation}
\gdep{X}{C} = -\hat{b} - \hat{\mathcal{L}}\,, \quad
\gdep{Y}{C}' = \hat{\mathcal{L}}\,, \quad
\gdep{Z}{C}_a' = \hat{\mathcal{X}}_a\,,
\end{equation}
while the \emph{total matter gauge} is based on the conditions
\begin{equation}
\ginv{b}{T} + \ginv{\mathcal{L}}{T} = \ginv{e}{T} = 0\,, \quad \ginv{f}{T}_a = 0\,,
\end{equation}
and thus satisfies
\begin{equation}
\gdep{X}{T} = -\hat{b} - \hat{\mathcal{L}}\,, \quad
\gdep{Y}{T} = -\hat{e}\,, \quad
\gdep{Z}{T}_a = -\hat{f}_a\,.
\end{equation}
While these gauges are familiar from other gravity theories, in the teleparallel class of gravity theories allows for further gauges, in which certain components of the teleparallel connection vanish, which allows to simplify the perturbed teleparallel field equations. For example, a gauge choice which turns out to be particularly useful in metric teleparallel gravity theories, and which has therefore been employed in~\cite{Bahamonde:2022ohm}, is in general (including also nonmetricity) defined by imposing the conditions
\begin{equation}
\ginv{j}{0} = \ginv{\sigma}{0} = 0\,, \quad
\ginv{z}{0}_a = 0\,,
\end{equation}
and leads to the transformation
\begin{equation}
\gdep{X}{0} = \hat{\sigma}' + \mathcal{Q}_3\hat{\sigma} - \hat{j}\,, \quad
\gdep{Y}{0} = -\hat{\sigma}\,, \quad
\gdep{Z}{0}_a = -\hat{z}_a\,.
\end{equation}
In the special case of symmetric teleparallel gravity, it turns out to be more convenient to use the homogeneous connection gauge, in which the connection perturbation~\eqref{eq:symconnpot} vanishes. This yields the condition
\begin{equation}
\ginv{\alpha}{H} = \ginv{\beta}{H} = 0\,, \quad
\ginv{\varkappa}{H}_a = 0\,,
\end{equation}
from which follows the gauge transformation
\begin{subequations}
\begin{align}
\gdep{X}{H}_{\perp} &= -\hat{\alpha}\,,\\
\gdep{X}{H}_{\parallel} &= -\hat{\beta}\,,\\
\gdep{Z}{H}_a &= -\hat{\varkappa}_a\,.
\end{align}
\end{subequations}
For the remainder of this article, we will exemplify the use of a convenient gauge choice.

\section{Energy-momentum-hypermomentum conservation}\label{sec:enmomhypcons}
As a first application of the formalism we present in this article, we study the conservation of energy-momentum and reduced hypermomentum in the teleparallel geometry. We first give an overview of the relevant conservation equations for the three types of teleparallel geometries in section~\ref{ssec:consgen}. We then discuss them in the context of cosmology, first for the homogeneous and isotropic background in section~\ref{ssec:consbkg} and finally for the perturbations in section~\ref{ssec:conspert}.

\subsection{Conservation equations}\label{ssec:consgen}
We start by briefly recalling the derivation of the energy-momentum-hypermomentum conservation from the general form of the matter action, whose variation we have written in the form~\eqref{eq:matactvar}. Under an infinitesimal diffeomorphism generated by a vector field \(V^{\mu}\), the variation of the action is then given by
\begin{equation}\label{eq:matactdiff}
\begin{split}
\delta_VS_{\text{m}} &= \int_M\left(\frac{1}{2}\Theta^{\mu\nu}(\mathcal{L}_Vg)_{\mu\nu} + H_{\mu}{}^{\nu\rho}(\mathcal{L}_V\Gamma)^{\mu}{}_{\nu\rho} + \Psi_I(\mathcal{L}_V\psi)^I\right)\sqrt{-g}\dd^4x\\
&= \int_M\left\{\Theta^{\mu\nu}\lc{\nabla}_{\mu}V_{\nu} + H_{\mu}{}^{\nu\rho}\left[\nabla_{\rho}\nabla_{\nu}V^{\mu} - \nabla_{\rho}(V^{\sigma}T^{\mu}{}_{\nu\sigma})\right] + \Psi_I\mathcal{L}_V\psi^I\right\}\sqrt{-g}\dd^4x\,.
\end{split}
\end{equation}
Here \(\Psi_I = 0\) are the matter field equations. Imposing that these hold (i.e., considering the variation on-shell), and performing integration by parts, the variation becomes
\begin{equation}
\delta_VS_{\text{m}} = -\int_M\left[\lc{\nabla}_{\nu}I_{\mu}{}^{\nu} - M^{\mu}{}_{\nu\rho}(I_{\mu}{}^{\nu} - \Theta_{\mu}{}^{\nu})\right]V^{\mu}\sqrt{-g}\dd^4x\,,
\end{equation}
making use of the definition~\eqref{eq:redhypmom}. It thus follows that the variation vanishes on-shell if and only if the energy-momentum and hypermomentum tensors satisfy the conservation law~\cite{Hohmann:2022mlc}
\begin{equation}\label{eq:enmomhypcons}
\lc{\nabla}_{\nu}I_{\mu}{}^{\nu} - M^{\rho}{}_{\nu\mu}(I_{\rho}{}^{\nu} - \Theta_{\rho}{}^{\nu}) = 0\,.
\end{equation}
In the derivation above we have not made any assumptions on the connection besides its flatness, which enters the form of the Lie derivative of the connection used in the variation~\eqref{eq:matactdiff}, and so the result is valid for all types of teleparallel geometries, including the general teleparallel geometry with both torsion and nonmetricity. If either of these quantities, we can replace the distortion tensor by either the contortion or disformation, and then further expand into the torsion or nonmetricity. In the metric teleparallel case, where the nonmetricity and hence the disformation \(L^{\mu}{}_{\nu\rho}\) vanishes, we can replace \(M^{\mu}{}_{\nu\rho}\) by \(K^{\mu}{}_{\nu\rho}\), and then write the energy-momentum-hypermomentum conservation as
\begin{equation}\label{eq:metenmomhypcons}
\nabla_{\nu}I_{\mu}{}^{\nu} + 2T^{\rho}{}_{\nu(\mu}I_{\rho)}{}^{\nu} = 0\,.
\end{equation}
For the symmetric teleparallel geometry, with vanishing torsion and thus vanishing contortion \(K^{\mu}{}_{\nu\rho}\), it is most convenient to work with densities, and now replacing the distortion \(M^{\mu}{}_{\nu\rho}\) by the disformation \(L^{\mu}{}_{\nu\rho}\) the conservation law can be written as
\begin{equation}\label{eq:symenmomhypcons}
\nabla_{\nu}\tilde{I}_{\mu}{}^{\nu} - \frac{1}{2}Q_{\mu\nu\rho}\tilde{\Theta}^{\nu\rho} = 0\,.
\end{equation}
Note that in the case of vanishing hypermomentum, \(H_{\mu}{}^{\nu\rho} = 0\), we have \(I_{\mu}{}^{\nu} = \Theta_{\mu}{}^{\nu}\), and thus all equations reduce to the familiar conservation equation
\begin{equation}
\lc{\nabla}_{\nu}\Theta_{\mu}{}^{\nu} = 0\,.
\end{equation}
We will study these equations for the cosmological background and its perturbations.

\subsection{Homogeneous and isotropic background}\label{ssec:consbkg}
We now discuss the energy-momentum-hypermomentum conservation laws derived in the previous section, where we assume that the metric, the connection, the energy-momentum and hypermomentum tensors take the homogeneous and isotropic background values given in section~\ref{sec:background}. In this case it follows that the only vanishing component of these equations, which transform as a covector, is the time component, since any spatial components must vanish identically due to the isotropy. In this case we find that the general conservation equation~\eqref{eq:enmomhypcons} becomes
\begin{equation}
\bar{\varrho}' + 3\mathcal{H}(\bar{\varrho} + \bar{\pi}) + \mathcal{Q}_1(\bar{\varrho} - \bar{\rho}) + 3\mathcal{Q}_2(\bar{\pi} - \bar{p}) = 0\,.
\end{equation}
Note in particular that any contributions from the torsion cancel, and that only two of the three time-dependent functions determining the nonmetricity enter the relation. It follows, and can also be shown by explicit calculation, that the same result is obtained also for the symmetric teleparallel case~\eqref{eq:symenmomhypcons}. In the metric teleparallel case~\eqref{eq:metenmomhypcons}, where the nonmetricity vanishes, the equation reduces to
\begin{equation}
\bar{\varrho}' + 3\mathcal{H}(\bar{\varrho} + \bar{\pi}) = 0\,.
\end{equation}
Finally, we remark that for vanishing hypermomentum, i.e., matter which does not couple to the teleparallel connection, so that \(\bar{\varrho} = \bar{\rho}\) and \(\bar{\pi} = \bar{p}\), all cases reduce to the common form
\begin{equation}
\bar{\rho}' + 3\mathcal{H}(\bar{\rho} + \bar{p}) = 0\,,
\end{equation}
which is the well-known cosmological energy-momentum conservation relation.

\subsection{Perturbations}\label{ssec:conspert}
We finally come to the energy-momentum-hypermomentum conservation at the linear perturbation order. To derive these relations, we make use of the background energy-momentum-hypermomentum conservation relations shown in the previous section, and impose that these are already satisfied. By making use of this assumption, we can express the perturbed energy-momentum-hypermomentum conservation in terms of gauge-invariant quantities. Here it turns out to be most convenient to express these through the Newtonian gauge defined in section~\ref{ssec:gaugeinv}. In particular, we find the time component
\begin{multline}
0 = \ginv{\mathcal{D}}{N}' + (3\mathcal{H} + \mathcal{Q}_1)\ginv{\mathcal{D}}{N} + 3(\mathcal{H} + \mathcal{Q}_2)\ginv{\mathcal{Q}}{N} - \mathcal{Q}_1\ginv{\mathcal{E}}{N} - 3\mathcal{Q}_2\ginv{\mathcal{P}}{N} + (\bar{\varrho} + \bar{\pi})\triangle\ginv{\mathcal{N}}{N}\\
+ (\bar{\varrho} - \bar{\rho})(\ginv{\phi}{N}' - \ginv{\varphi}{N}') - 3(\bar{\varrho} + \bar{p})\ginv{\psi}{N}' + (\bar{\pi} - \bar{p})(\triangle\ginv{\sigma}{N}' - 3\ginv{\varpi}{N}')\,,
\end{multline}
while the spatial component decomposes into a scalar total divergence
\begin{multline}
0 = [(\bar{\varrho} + \bar{\pi})\ginv{\mathcal{M}}{N}]' + (\bar{\varrho} + \bar{\pi})\left[(3\mathcal{H} + \mathcal{T}_1 - \mathcal{Q}_2)\ginv{\mathcal{M}}{N}_a + (\mathcal{H} - \mathcal{T}_1 + \mathcal{Q}_2 - \mathcal{Q}_3)\ginv{\mathcal{N}}{N}\right] + \frac{2}{3}\triangle\ginv{\mathcal{A}}{N} + 2u^2\ginv{\mathcal{A}}{N} + (\bar{\rho} + \bar{p})\mathcal{Q}_3\ginv{\mathcal{L}}{N} + 2\mathcal{T}_2\ginv{\mathcal{I}}{N} + \ginv{\mathcal{Q}}{N}\\
- \frac{1}{2}(\bar{\varrho} - \bar{\rho} + \bar{\pi} - \bar{p})\left[\mathcal{Q}_3\ginv{y}{N} - (2\mathcal{H} - 2\mathcal{T}_1 + 2\mathcal{Q}_2 - \mathcal{Q}_3)\ginv{j}{N}\right] - (\bar{\pi} - \bar{p})(\triangle\ginv{\sigma}{N} - 3\ginv{\varpi}{N} + 3\ginv{\psi}{N}) + (\bar{\varrho} - \bar{\rho})\ginv{\varphi}{N} + (\bar{\rho} + \bar{\pi})\ginv{\phi}{N}
\end{multline}
and a divergence-free vector part
\begin{multline}
0 = [(\bar{\varrho} + \bar{\pi})\ginv{\mathcal{Y}}{N}_a]' + (\bar{\varrho} + \bar{\pi})\left[(3\mathcal{H} + \mathcal{T}_1 - \mathcal{Q}_2)\ginv{\mathcal{Y}}{N}_a + (\mathcal{H} - \mathcal{T}_1 + \mathcal{Q}_2 - \mathcal{Q}_3)\ginv{\mathcal{Z}}{N}_a\right] + \frac{1}{2}\triangle\ginv{\mathcal{B}}{N}_a + u^2\ginv{\mathcal{B}}{N}_a\\
+ (\bar{\rho} + \bar{p})\mathcal{Q}_3\ginv{\mathcal{X}}{N}_a + 2\mathcal{T}_2\ginv{\mathcal{J}}{N}_a + \upsilon_{abc}\dd^b\ginv{\mathcal{J}}{N}^c - \frac{1}{2}(\bar{\varrho} - \bar{\rho} + \bar{\pi} - \bar{p})\left[\mathcal{Q}_3\ginv{v}{N}_a - (2\mathcal{H} - 2\mathcal{T}_1 + 2\mathcal{Q}_2 - \mathcal{Q}_3)\ginv{c}{N}_a\right]\,.
\end{multline}
As for the background quantities discussed in section~\ref{ssec:consbkg}, these equations constrain the time evolution of some of the perturbations as functions of the others. However, they do not determine all components, and so the remaining components must be determined from the given matter model. Yet any matter model which satisfies diffeomorphism invariance will obey the energy-momentum-hypermomentum conservation equations given above.

\section{Example: tensor perturbations in $f(X)$ type theories}\label{sec:example}
In order to illustrate the formalism presented in this article, we derive the field equations for the tensorial perturbations in a number of gravity theories, which cover all possible flavors of teleparallel geometries. Each of these classes of theories is defined by a gravitational Lagrangian, which is a free function \(f\) of a scalar, which agrees with the Ricci scalar of the Levi-Civita connection up to a boundary term. In the metric teleparallel geometry, this leads to the class of \(f(T)\) theories, which we discuss in section~\ref{ssec:fT}. Analogously, in the symmetric teleparallel geometry we have the class of \(f(Q)\) theories, discussed in section~\ref{ssec:fQ}. Finally, the general teleparallel geometry yields the class of \(f(G)\) theories, which we study in section~\ref{ssec:fG}.

\subsection{$f(T)$ gravity}\label{ssec:fT}
For the first example, we study the metric teleparallel geometry, in which the nonmetricity~\eqref{eq:nonmetricity} is imposed to vanish. In this case, also the disformation~\eqref{eq:disformation} vanishes, and it follows from the connection decomposition~\eqref{eq:conndec} that the distortion and contortion agree. Hence, using the curvature relation~\eqref{eq:curvdec}, we can write the Ricci scalar of the Levi-Civita connection as
\begin{equation}
\lc{R} = -T + B_T\,,
\end{equation}
where we have introduced the terms
\begin{equation}
T = 2K^{\mu}{}_{\tau[\mu}K^{\tau\nu}{}_{\nu]}\,, \quad
B_T = 2\lc{\nabla}_{\mu}K^{[\nu\mu]}{}_{\nu}\,.
\end{equation}
Clearly, \(B_T\) is a total divergence, and would thus take the role of a boundary term under an action integral, which does not contribute to the field equations. Hence, replacing \(\lc{R}\) by \(-T\) in the Einstein-Hilbert action yields the same metric field equations as the original action. Here we consider a modification of this equivalent action, which takes the form~\cite{Bengochea:2008gz,Linder:2010py,Krssak:2015oua}
\begin{equation}
S_{\text{g}} = -\frac{1}{2\kappa^2}\int_Mf(T)\sqrt{-g}\dd^4x\,,
\end{equation}
with a free function \(f\). Since we are interested in the dynamics of the tensor perturbations only, it is sufficient to study the metric field equation, obtained from the variation of this action with respect to the metric tensor, together with the matter contribution~\eqref{eq:matactvar}. The perturbed field equations have been derived for a spatially flat background in~\cite{Golovnev:2018wbh} and for all cosmologically symmetric branches in~\cite{Bahamonde:2022ohm}, and so we briefly review the corresponding equations here. Considering first the case \(u \neq 0\), we have the branch~\eqref{eq:cosmogaxi} which exhibits axial torsion. In the metric teleparallel case, we set \(\mathcal{T}_1 = \mathcal{H}\) and \(\mathcal{T}_2 = u\), and obtain
\begin{equation}\label{eq:tensTaxi}
2\kappa^2A^2\hat{\mathcal{T}}_{ab} = f_T\left(\triangle\hat{q}_{ab} - 2u^2\hat{q}_{ab} - 2\mathcal{H}\hat{q}_{ab}' - \hat{q}_{ab}''\right) + 12\frac{f_{TT}}{A^2}\mathcal{H}(\mathcal{H}^2 - u^2 - \mathcal{H}')\hat{q}_{ab}'\,.
\end{equation}
Note that we could equivalently have set \(\mathcal{T}_2 = -u\). Similarly, the branch~\eqref{eq:cosmogvec} is achieved for \(\mathcal{T}_1 = \mathcal{H} + iu\) and \(\mathcal{T}_2 = 0\), and yields
\begin{equation}\label{eq:tensTvec}
2\kappa^2A^2\hat{\mathcal{T}}_{ab} = f_T\left(\triangle\hat{q}_{ab} - 2u^2\hat{q}_{ab} - 2\mathcal{H}\hat{q}_{ab}' - \hat{q}_{ab}''\right) + 12\frac{f_{TT}}{A^2}(\mathcal{H} + iu)[\mathcal{H}(\mathcal{H} + iu) - \mathcal{H}']\left(\hat{q}_{ab}' - iu\hat{q}_{ab}\right)\,.
\end{equation}
Again, changing the sign of \(u\) in does not give a qualitative difference. Finally, for \(u = 0\), both branches give the limit \(\mathcal{T}_1 = \mathcal{H}\) and \(\mathcal{T}_2 = 0\), for which the tensor equations become
\begin{equation}\label{eq:tensTflat}
2\kappa^2A^2\hat{\mathcal{T}}_{ab} = f_T\left(\triangle\hat{q}_{ab} - 2\mathcal{H}\hat{q}_{ab}' - \hat{q}_{ab}''\right) + 12\frac{f_{TT}}{A^2}\mathcal{H}(\mathcal{H}^2 - \mathcal{H}')\hat{q}_{ab}'\,.
\end{equation}
We see that apart from the factor \(f_T\), which acts as an effective gravitational constant, these equations reproduce the known result from general relativity, with gravitational waves propagating at the speed of light, while the additional terms proportional to \(f_{TT}\) contribute to the Hubble friction and curvature terms only.

\subsection{$f(Q)$ gravity}\label{ssec:fQ}
We then continue with the symmetric teleparallel geometry, where we impose vanishing torsion~\eqref{eq:torsion}, and hence vanishing contortion~\eqref{eq:contortion}. From the curvature decomposition~\eqref{eq:curvdec} and the connection decomposition~\eqref{eq:conndec} then follows that we can write the Ricci scalar in the form
\begin{equation}
\lc{R} = -Q + B_Q\,,
\end{equation}
where the two terms on the right-hand side are given by
\begin{equation}
Q = 2L^{\mu}{}_{\tau[\mu}L^{\tau\nu}{}_{\nu]}\,, \quad
B_Q = 2\lc{\nabla}_{\mu}L^{[\nu\mu]}{}_{\nu}\,.
\end{equation}
Also here the second term \(B_Q\) is a total divergence, and so its contribution to the Einstein-Hilbert action is a pure boundary term, which can be neglected without changing the metric field equations, when we replace \(\lc{R}\) by \(-Q\) in the action. Here we generalize the action to~\cite{BeltranJimenez:2017tkd}
\begin{equation}
S_{\text{g}} = -\frac{1}{2\kappa^2}\int_Mf(Q)\sqrt{-g}\dd^4x\,.
\end{equation}
Also in this case the only relevant field equation for the tensor perturbations is obtained by variation of the action with respect to the metric, with the matter contribution originating from the variation~\eqref{eq:matactvar}. We study these tensor perturbation equations for the different cosmologically symmetric background geometries derived in~\cite{Hohmann:2021ast,DAmbrosio:2021pnd}. For the spatially curved case \(u \neq 0\), the branch~\eqref{eq:cosmogaxi} mandates a non-vanishing axial torsion, and so this branch is not present in the symmetric teleparallel geometry. Hence, we are left with the only spatially curved branch~\eqref{eq:cosmogvec}, where the nonmetricity can be parametrized as
\begin{equation}
\mathcal{Q}_1 = \mathcal{H} - \mathcal{K} - \frac{\dot{\mathcal{K}}}{\mathcal{K}}\,, \quad
\mathcal{Q}_2 = \mathcal{K} - \mathcal{H}\,, \quad
\mathcal{Q}_3 = \mathcal{K} + \frac{u^2}{\mathcal{K}}\,.
\end{equation}
The field equation for the tensor perturbations then becomes
\begin{multline}\label{eq:tensQcurv}
2\kappa^2A^2\hat{\mathcal{T}}_{ab} = f_Q\left(\triangle\hat{q}_{ab} - 2u^2\hat{q}_{ab} - 2\mathcal{H}\hat{q}_{ab}' - \hat{q}_{ab}''\right)\\
+ 3\frac{f_{QQ}}{A^2\mathcal{K}^4}\left\{4\mathcal{H}\mathcal{K}(\mathcal{H} - \mathcal{K})(\mathcal{H}\mathcal{K} + u^2) + 2\mathcal{K}[(\mathcal{K}^2 - 2\mathcal{H}\mathcal{K} - u^2)\mathcal{H}'+ (u^2 + \mathcal{K}^2)\mathcal{K}'']\right\}(\mathcal{K}\hat{q}_{ab}' - 2u^2\hat{q}_{ab})\,.
\end{multline}
We see that similarly to the metric teleparallel case we obtain a contribution to the Hubble friction and curvature terms, while the speed of propagation remains equal to the speed of light. We are left with the three spatially flat branches. In the case~\eqref{eq:cosmogcoin}, we can set
\begin{equation}
\mathcal{Q}_1 = \mathcal{K}\,, \quad
\mathcal{Q}_2 = -\mathcal{H}\,, \quad
\mathcal{Q}_3 = 0\,.
\end{equation}
The tensor field equation then becomes
\begin{equation}\label{eq:tensQcoin}
2\kappa^2A^2\hat{\mathcal{T}}_{ab} = f_Q\left(\triangle\hat{q}_{ab} - 2\mathcal{H}\hat{q}_{ab}' - \hat{q}_{ab}''\right) + 12\frac{f_{QQ}}{A^2}\mathcal{H}(\mathcal{H}^2 - \mathcal{H}')\hat{q}_{ab}'\,,
\end{equation}
and thus qualitatively agrees with the metric teleparallel case~\eqref{eq:tensTvec}, with a modified Hubble friction only; note, however, that the different background evolution in these theories still enters through the derivatives of the function \(f\) appearing in this equation. For the branch~\eqref{eq:cosmogconf}, we choose the parametrization
\begin{equation}
\mathcal{Q}_1 = \mathcal{H} + \frac{\mathcal{K}'}{\mathcal{K}}\,, \quad
\mathcal{Q}_2 = -\mathcal{H}\,, \quad
\mathcal{Q}_3 = \mathcal{K}\,,
\end{equation}
and find that the tensor perturbation equations
\begin{equation}\label{eq:tensQconf}
2\kappa^2A^2\hat{\mathcal{T}}_{ab} = f_Q\left(\triangle\hat{q}_{ab} - 2\mathcal{H}\hat{q}_{ab}' - \hat{q}_{ab}''\right) + 3\frac{f_{QQ}}{A^2}[4\mathcal{H}^2(\mathcal{H} - \mathcal{K}) - 2(2\mathcal{H} - \mathcal{K})\mathcal{K}' + \mathcal{K}'']\hat{q}_{ab}'\,,
\end{equation}
where also here only the Hubble friction receives a modification. Finally, in the branch~\eqref{eq:cosmogpara}, we can set
\begin{equation}
\mathcal{Q}_1 = \mathcal{H} - \mathcal{K} - \frac{\mathcal{K}'}{\mathcal{K}}\,, \quad
\mathcal{Q}_2 = \mathcal{K} - \mathcal{H}\,, \quad
\mathcal{Q}_3 = \mathcal{K}\,,
\end{equation}
which yields the equation
\begin{equation}\label{eq:tensQpara}
2\kappa^2A^2\hat{\mathcal{T}}_{ab} = f_Q\left(\triangle\hat{q}_{ab} - 2\mathcal{H}\hat{q}_{ab}' - \hat{q}_{ab}''\right) + 3\frac{f_{QQ}}{A^2}[4\mathcal{H}^2(\mathcal{H} + \mathcal{K}) - 2(2\mathcal{H} + \mathcal{K})\mathcal{K}' - \mathcal{K}''](\hat{q}_{ab}' - 2\mathcal{K}\hat{q}_{ab})
\end{equation}
for the tensor perturbations. Again we see modified Hubble friction and curvature terms, while the speed of propagation is unchanged.

\subsection{$f(G)$ gravity}\label{ssec:fG}
We finally come to the general teleparallel geometry, in which both torsion and nonmetricity may be non-vanishing. In this case the curvature decomposition~\eqref{eq:curvdec} immediately yields
\begin{equation}
\lc{R} = -G + B_G\,,
\end{equation}
where we defined the terms
\begin{equation}
G = 2M^{\mu}{}_{\tau[\mu}M^{\tau\nu}{}_{\nu]}\,, \quad
B_G = 2\lc{\nabla}_{\mu}M^{[\nu\mu]}{}_{\nu}\,.
\end{equation}
As discussed in the metric and symmetric teleparallel cases, replacing \(\lc{R}\) by \(-G\) in the Einstein-Hilbert action does not change the metric field equations, since \(B_G\) is a boundary term and hence does not contribute. Applying an arbitrary function, such that the action becomes~\cite{Hohmann:2022mlc,Heisenberg:2022mbo}
\begin{equation}
S_{\text{g}} = -\frac{1}{2\kappa^2}\int_Mf(G)\sqrt{-g}\dd^4x\,,
\end{equation}
however, breaks this equivalence. In the following, we will study the dynamics of tensor perturbations around the different cosmological background geometries in this class of theories. Note that in contrast to the previously discussed cases, the general teleparallel geometry allows for a second tensor perturbation \(\hat{u}_{ab}\) originating from the teleparallel connection, in addition to the metric perturbation \(\hat{q}_{ab}\), and also the field equation obtained by variation with respect to the connection contains a tensor component. However, in the particular case of \(f(G)\) gravity, it turns out that \(\hat{u}_{ab}\) does not enter the field equations, and that the tensor part of the connection equations is satisfied identically. One therefore finds a single equation for the perturbation \(\hat{q}_{ab}\), which we now show for the different branches of background geometries derived in~\cite{Heisenberg:2022mbo}. For the case~\eqref{eq:cosmogaxi}, we can parametrize the cosmologically symmetric teleparallel connection as
\begin{equation}
\mathcal{T}_1 = \mathcal{H} + \mathcal{K}\,, \quad
\mathcal{T}_2 = u\,, \quad
\mathcal{Q}_1 = \mathcal{L}\,, \quad
\mathcal{Q}_2 = \mathcal{K}\,, \quad
\mathcal{Q}_3 = 0\,.
\end{equation}
In this case the dynamics of the tensor perturbations is governed by the equation
\begin{equation}\label{eq:tensGaxi}
2\kappa^2A^2\hat{\mathcal{T}}_{ab} = f_G\left(\triangle\hat{q}_{ab} - 2u^2\hat{q}_{ab} - 2\mathcal{H}\hat{q}_{ab}' - \hat{q}_{ab}''\right) + 12\frac{f_{GG}}{A^2}\mathcal{H}(\mathcal{H}^2 - u^2 - \mathcal{H}')\hat{q}_{ab}'\,,
\end{equation}
and so formally agrees with the dynamics~\eqref{eq:tensTaxi} in the metric teleparallel case, except for the background value of the function \(f\), which now also depends on the two additional dynamical functions \(\mathcal{K}\) and \(\mathcal{L}\). For the remaining spatially curved branch~\eqref{eq:cosmogvec}, we choose the parametrization
\begin{equation}
\mathcal{T}_1 = \mathcal{H} + \mathcal{K} - \mathcal{L}\,, \quad
\mathcal{T}_2 = 0\,, \quad
\mathcal{Q}_1 = -\mathcal{K} - \frac{\mathcal{L}'}{\mathcal{L}}\,, \quad
\mathcal{Q}_2 = \mathcal{K}\,, \quad
\mathcal{Q}_3 = \mathcal{L} + \frac{u^2}{\mathcal{L}}\,.
\end{equation}
Here we find the tensor field equation
\begin{equation}\label{eq:tensGvec}
2\kappa^2A^2\hat{\mathcal{T}}_{ab} = f_G\left(\triangle\hat{q}_{ab} - 2u^2\hat{q}_{ab} - 2\mathcal{H}\hat{q}_{ab}' - \hat{q}_{ab}''\right)\,,
\end{equation}
which agrees with the known result from general relativity, except for the factor \(f_G\) modifying the effective gravitational constant. The propagation of gravitational waves in vacuum, however, is unaltered by this modification. We then continue with the spatially flat cases. For the branch~\eqref{eq:cosmogcoin}, we can choose the parametrization
\begin{equation}
\mathcal{T}_1 = \mathcal{H} + \mathcal{K}\,, \quad
\mathcal{T}_2 = 0\,, \quad
\mathcal{Q}_1 = \mathcal{L}\,, \quad
\mathcal{Q}_2 = \mathcal{K}\,, \quad
\mathcal{Q}_3 = 0\,,
\end{equation}
and find the tensor field equation
\begin{equation}\label{eq:tensGcoin}
2\kappa^2A^2\hat{\mathcal{T}}_{ab} = f_G\left(\triangle\hat{q}_{ab} - 2\mathcal{H}\hat{q}_{ab}' - \hat{q}_{ab}''\right) + 12\frac{f_{GG}}{A^2}\mathcal{H}(\mathcal{H}^2 - \mathcal{H}')\hat{q}_{ab}'\,,
\end{equation}
which represents the limit \(u \to 0\) of the previously found equation~\eqref{eq:tensGaxi}. Similarly, considering the branch~\eqref{eq:cosmogconf} and the parametrization
\begin{equation}
\mathcal{T}_1 = \mathcal{H} + \mathcal{K} - \mathcal{L}\,, \quad
\mathcal{T}_2 = 0\,, \quad
\mathcal{Q}_1 = -\mathcal{K} - \frac{\mathcal{L}'}{\mathcal{L}}\,, \quad
\mathcal{Q}_2 = \mathcal{K}\,, \quad
\mathcal{Q}_3 = \mathcal{L}\,,
\end{equation}
the dynamics of tensor perturbations is governed by the equation
\begin{equation}\label{eq:tensGflat}
2\kappa^2A^2\hat{\mathcal{T}}_{ab} = f_G\left(\triangle\hat{q}_{ab} - 2\mathcal{H}\hat{q}_{ab}' - \hat{q}_{ab}''\right)\,.
\end{equation}
Also here we see only a modification of the effective gravitational constant in comparison to general relativity, as the equation is obtained from~\eqref{eq:tensGvec} in the limit \(u \to 0\). The same holds for the branch~\eqref{eq:cosmogpara} parametrized by
\begin{equation}
\mathcal{T}_1 = \mathcal{H} + \mathcal{K}\,, \quad
\mathcal{T}_2 = 0\,, \quad
\mathcal{Q}_1 = -\mathcal{K} + \frac{\mathcal{L}'}{\mathcal{L}}\,, \quad
\mathcal{Q}_2 = \mathcal{K}\,, \quad
\mathcal{Q}_3 = \mathcal{L}\,,
\end{equation}
for which we find the same equation~\eqref{eq:tensGflat}. Hence, we find that the tensor perturbations in the \(f(G)\) class of gravity theories are formally described by the same set of equations as in general relativity and in \(f(T)\) gravity, and observational differences can appear only through a modified background evolution influencing the cosmological background value of the function \(f\) and its derivatives in the tensor field equations.

\section{Conclusion}\label{sec:conclusion}
We have studied the most general perturbations of general, metric and symmetric teleparallel geometries around a cosmologically symmetric, i.e., homogeneous and isotropic background. In particular, we have decomposed the perturbations of the metric and the teleparallel connection, the matter variables and the general form of the gravitational field equations into irreducible components, and studied their behavior under gauge transformations. Making use of the latter, we have shown how gauge-invariant quantities can be constructed from these components. To show their use, we have derived the energy-momentum-hypermomentum conservation equations as a linear perturbation around the cosmologically symmetric background.

As a further example, we have studied the propagation of tensor perturbations in the $f(T)$, $f(Q)$ and $f(G)$ classes of gravity theories. Our findings have shown that these receive only minimal corrections compared to the well-known case of general relativity, affecting the effective mass and Hubble friction terms. In particular, we find that for all of these theories only two tensor modes propagate, despite the fact that the general teleparallel geometry perturbation contains another tensor mode.

While in our work we have focused on the geometry of perturbations and their gauge transformations, we leave the study of the dynamics of these perturbations in different teleparallel gravity theories for future work. As an example, we present study the perturbative degrees of freedom in the $f(T)$, $f(Q)$ and $f(G)$ classes of gravity theories with a minimally coupled scalar field around cosmological backgrounds in~\cite{Heisenberg:2023wgk}. As another line of research, we plan to generalize our work to more general cosmological backgrounds, which are only homogeneous, but not isotropic, such as Bianchi cosmologies~\cite{Hohmann:2023sto}.

\begin{acknowledgments}
LH is supported by funding from the European Research Council (ERC) under the European Unions Horizon 2020 research and innovation programme grant agreement No 801781. LH further acknowledges support from the Deutsche Forschungsgemeinschaft (DFG, German Research Foundation) under Germany's Excellence Strategy EXC 2181/1 - 390900948 (the Heidelberg STRUCTURES Excellence Cluster). MH gratefully acknowledges the full Financial support by the Estonian Research Council through the Personal Research Funding project PRG356. The authors acknowledge networking support by the COST Actions CA18108 and CA21136.
\end{acknowledgments}

\bibliography{telecosmo}

\end{document}